\documentclass[12pt,UTF8]{article}
\usepackage{mathrsfs}
\usepackage{graphicx,setspace,lscape,longtable,float}
\usepackage{subfig}
\usepackage{stmaryrd}
\usepackage{multirow}
\usepackage{array}
\usepackage{tikz}
\usepackage{bm}
\usepackage{booktabs}
\usepackage[round]{natbib}
\usepackage{epsfig,graphicx}
\usepackage{mathrsfs,amsmath,amsthm,amssymb,color}
\usepackage{enumerate}
\usepackage{cases}
\usepackage{xr}
\usepackage{authblk}
\usepackage[T1]{fontenc}
\usepackage{rotating}
\usepackage[hang,flushmargin,perpage,symbol]{footmisc}
\usepackage{algorithm}

\setfnsymbol{wiley}

\newtheorem{theorem}{Theorem}
\newtheorem{lemma}{Lemma}
\newtheorem{assumption}{Assumption}
\newtheorem{proposition}{Proposition}
\newtheorem{corollary}{Corollary}
\newtheorem{remark}{Remark}
\newtheorem{definition}{Definition}

\def\beq{\begin{equation}}
\def\eeq{\end{equation}}
\def\beqr{\begin{eqnarray}}
\def\eeqr{\end{eqnarray}}
\def\bet{\begin{theorem}}
\def\eet{\end{theorem}}
\def\bel{\begin{lemma}}
\def\eel{\end{lemma}}
\def\bep{\begin{proposition}}
\def\eep{\end{proposition}}

\def\mR{\mathbb{R}}

\numberwithin{equation}{section}

\begin{document}
\def\spacingset#1{\renewcommand{\baselinestretch}%
	{#1}\small\normalsize} \spacingset{1}
	
\title{\bf   Angle-based hierarchical classification using exact label embedding}
\author[1]{Yiwei Fan}
\author[1]{Xiaoling Lu}
\author[2]{Yufeng Liu}
\author[3]{Junlong Zhao$^{*,}$}
\affil[1]{\it\small  Renmin University of China,  China}
\affil[2]{\it\small University of North Carolina at Chapel Hill, USA}
\affil[3]{\it\small Beijing Normal University, China}
%\author{}
%\renewcommand\Authands{ and }
\date{}
\maketitle

\footnote{$^*$Corresponding author. E-mail address: zhaojunlong928@126.com. Junlong  Zhao was supported by National Natural Science Foundation of China  (No.11471030). Xiaoling Lu was supported by National Natural Science Foundation of China (No.61472475).}

\begin{abstract}
\spacingset{1.5}
Hierarchical classification problems are commonly seen in practice. However, most existing methods do not fully utilize the hierarchical information among class labels. In this paper, a novel label embedding approach is proposed, which keeps the hierarchy of labels exactly, and reduces the complexity of the hypothesis space significantly. Based on the newly proposed label embedding approach, a new angle-based classifier is developed for hierarchical classification.  Moreover, to handle massive data, a new (weighted) linear loss is designed, which has a closed form solution and is computationally efficient. Theoretical properties of the new method are established and intensive numerical comparisons with other methods are  conducted. Both simulations and applications in document categorization demonstrate the advantages of the proposed method.
\end{abstract}

\noindent%
\vfill
\newpage
\spacingset{1.7} % DON'T change the spacing!

\section{Introduction}
{
Hierarchical classification problems are  commonly encountered in many scientific  fields \citep{Silla:2011}, including but not limited to image classification \citep{Akata:2015,Akata:2016}, text categorization \citep{Koller:1997},  protein function prediction \citep{Vens:2008}, music genre classification \citep{DeCoro:2007, Silla:2009b}, and online commerce \citep{Chen:2013}.
In hierarchical classification, the hierarchy of the classes can be pre-defined by a graph, where each node stands for a class,   and a directed edge from the node $\nu$ to the node $\nu'$ means that if an instance is assigned to $\nu'$, then it must be assigned to $\nu$ first. We call $\nu$ a parent of $\nu'$, and $\nu'$ a child of $\nu$.
%An ancestor is a parent or (recursively) the parent of an ancestor. Siblings are nodes that sharing the same parent(s).
A node without any child is referred to a leaf, and it does not necessarily locate at the last layer. If each node has at most one parent, then the graph is of a tree structure; otherwise, we call it a Directed Acyclic Graph (DAG). An illustrative example of a tree structure is shown in Figure 1.
In this article, we focus on tree structures. Moreover, we assume that each node either is a leaf or has at least two children, and that an instance to be classified belongs to at most one node at any layer in the hierarchy, namely single-labeled.

For hierarchical classification, the simplest approach is to apply a flat classifier, which predicts only the leaf nodes, completely ignoring the hierarchy \citep{Hayete:2005, Barbedo:2006}. Another popular approach is to sequentially train a multicategory classifier locally at each parent node \citep{davies2007hierarchical}, or train a binary classifier at each node \citep{cesa2006hierarchical}. The classifier may suffer from a small training sample and be suboptimal. Besides, to incorporate the hierarchical structure among nodes in learning classification rules, various other methods have been developed during the past several decades including
%training one binary classifier for each node \citep{Eisner:2005,Fagni:2007},
imposing inequality constraints directly \citep{Wang:2009}, designing regularized loss functions \citep{Gopal:2013}, and considering cost-sensitive learning \citep{Fan:2015, Charuvaka:2015}.  A detailed survey of hierarchical classification can be found in \citet{Silla:2011}.
%To reduce the computation time,  some implicit approaches  are developed recent years  such as the recursive regularization approach of  Gopal and Yang (2013) and the HierCost (Charuvaka and Rangwala, 2015), borrowing some  ideas of cost-sensitive learning and multi-task learning.
%These methods   usually depend on specifically designed loss function or  classifier to incorporate the hierarchical structure.
%Compared with the methods imposing  inequality constraints,  these  approaches  avoid the inequality constraints and have  significant improvement in computation.
%However, when the sample size is very large, the computation is still heavy.

Besides the methods mentioned above, several methods on label embedding have been developed for both multicategory classification \citep{Lange:2008, Wu:2010, Wu:2012, Zhang:2014} and hierarchical classification \citep{Cai:2004, Tsochantaridis:2005, Bengio:2010}, though some of them did not use this particular term. Label embedding  aims to map nodes into a set of points in the Euclidean space, such that the Euclidean distance between these points  mimics the dissimilarity between the nodes as much as possible. Such a method has obvious  advantages in computation, as the classification problem can be naturally transformed into a regression task.
It has been proven useful in many application domains such as image classification \citep{Akata:2016, Chollet:2016} and text categorization \citep{Weinberger:2009}.

For hierarchical classification, there are two crucial factors for the success of label embedding, the hierarchy of the embedded points and the dimension of
the embedded Euclidean space.
A desired embedding approach is to embed nodes into points in a low-dimensional space while keeping the hierarchy.
Suppose there are $q$ nodes totally excluding the root. The classical approach  maps each node into a $q$-dimensional vector \citep{Cai:2004,Tsochantaridis:2005}.
As shown in Section 2, the Euclidean distance between these vectors cannot mimic the dissimilarity between the nodes properly, and thus they do not maintain the hierarchy well. In addition, the embedded space has the dimension $q$, which leads to a complex hypothesis space.
To reduce the dimension of the embedded space, some existing papers developed approximated embedding \citep{Weinberger:2009,Bengio:2010}. However, these approximated  embedding approaches cannot keep the hierarchy exactly. It is desirable to propose an embedding method such that the embedded points keep the hierarchy exactly as well as locate in a low-dimensional Euclidean space.

Motivated by the existing work, we develop a label embedding method that keeps the hierarchy exactly, i.e. it satisfies two basic properties of  dissimilarities between nodes on the hierarchical tree. Surprisingly,  the dimension of the embedded space is only $n_{\mathrm{leaf}}-1$, where   $n_{\mathrm{leaf}}$ denotes  the number of leaf nodes, much smaller than $q$, the number of nodes excluding the root, especially when the tree is complicated.
In addition, note that this dimension is  exactly the  same  as the  one required for  the  multicategory classification  on leaf nodes by the label embedding approach of \citet{Lange:2008} and \cite{Zhang:2014}.
This observation   sheds  light on the long-standing
phenomenon that flat classifiers which ignore the hierarchy, can still be competitive, compared with some hierarchical classifiers  \citep{Babbar:2013, Hoyoux:2016}.
Flat classifiers directly applying to leaf nodes involve a lower-dimensional hypothesis space, while some hierarchical classifiers utilizing the hierarchical information involve a much higher-dimensional hypothesis space.
In this sense, our embedding method takes the advantages of both aspects,
keeping the hierarchy exactly and reducing the complexity of the hypothesis space (or equivalently the number of unknown parameters) simultaneously.
Based on our embedded points, we then extend the angle-based method  \citep{Zhang:2014} to the hierarchical case.

There are several  key contributions in this paper. Firstly, we address that an ideal dissimilarity measurement between nodes on the hierarchical tree should satisfy two basic properties, called hierarchical and symmetric (H.S.) properties. Then we define a novel dissimilarity measurement that satisfies the properties. Secondly, we develop an exact label embedding procedure to construct points in  a low-dimensional Euclidean space. Thirdly, we propose an angle-based method for hierarchical classification and establish some statistical properties. The convergence rate of the proposed method has advantages over existing approaches.
Fourthly, we design a (weighted) linear loss function, under which  the estimator can be derived in a closed form without complicated optimization. It is particularly useful when the tree is complex or the sample size is large, which is the case in big data analyses.

The remaining of this article is organized as follows. In Section 2, we state H.S. properties and define a novel dissimilarity measurement.  In Section 3, we develop an exact label embedding procedure. In Section 4, we propose the angle-based method and the linear loss functions.  Some theoretical properties of the estimator including Fisher consistency and asymptotic results on generalization errors are  established in Section 5.  Simulations and real data analyses are presented in Section 6. Finally, we make discussions in Section 7.

Before  proceeding, we introduce some notations used in the paper. For any positive integers $m$ and $i$ with  $i\le m$, let $\bm{e}_i=(0,\cdots,0,1,0,\cdots,0)^{\top}\in \mR^m$  with the $i$-th coordinate being 1 and others being 0. For any vector $\bm{u}=(u_1,\cdots, u_m)^{\top}\in \mR^m$, $\|\bm{u}\|$ denotes the $l_2$ norm, and  $\bm{u}^{(\widetilde{m})}$ denotes the subvector consisting of the first $\widetilde{m}$ coordinates of $\bm{u}$ with $\widetilde{m}\leq m$. For any set $S$, $|S|$ denotes the cardinality of $S$.

\section{Dissimilarity between classes}

\subsection{Hierarchical and symmetric (H.S.) properties}

We first introduce some notations and definitions.
The ancestor of a node is its parent or recursively the parent of an ancestor. The offspring of a node is referred to its  child or the child of an offspring.
Siblings are nodes sharing the same parent. For a node, denote its parent, children, ancestors, offsprings and siblings respectively as Par($\cdot$), Chi($\cdot$), Anc($\cdot$), Off($\cdot$), and Sib($\cdot$).
For a tree structure, assume it has $k$ layers in total. Denote $C_1$ as the node at the first layer which is the root, and  $C_{1,j_{2}}$ as the child of $C_1$ with index $j_{2}=1,2,\cdots , N_{1}$ at the second layer from left to right,  where $N_{1}$ is the number of children for the node $C_{1}$.
In general, for $3\leq m \leq k$, $C_{1,j_{2}, \cdots, j_{m-1}, j_{m}}$ denotes the child of $C_{1,j_{2}, \cdots, j_{m-1}}$ with index $j_{m}=1,2,\cdots , N_{1,j_{2}, \cdots , j_{m-1}}$ at the $m$-th layer from left to right, where $N_{1,j_{2}, \cdots, j_{m-1}}$ denotes the number of children for $C_{1,j_{2}, \cdots, j_{m-1}}$. Denote the collection of all nodes as
$\mathcal{C}=\{C_1\}\cup\{C_{j_1,j_2,\cdots, j_s}: j_1\equiv1, j_s=1,\cdots, N_{j_1,j_2,\cdots, j_{s-1}}, s=2,\cdots,k\  \}.$
\begin{figure}[h]
	\centering
	\includegraphics[width=0.85\textwidth]{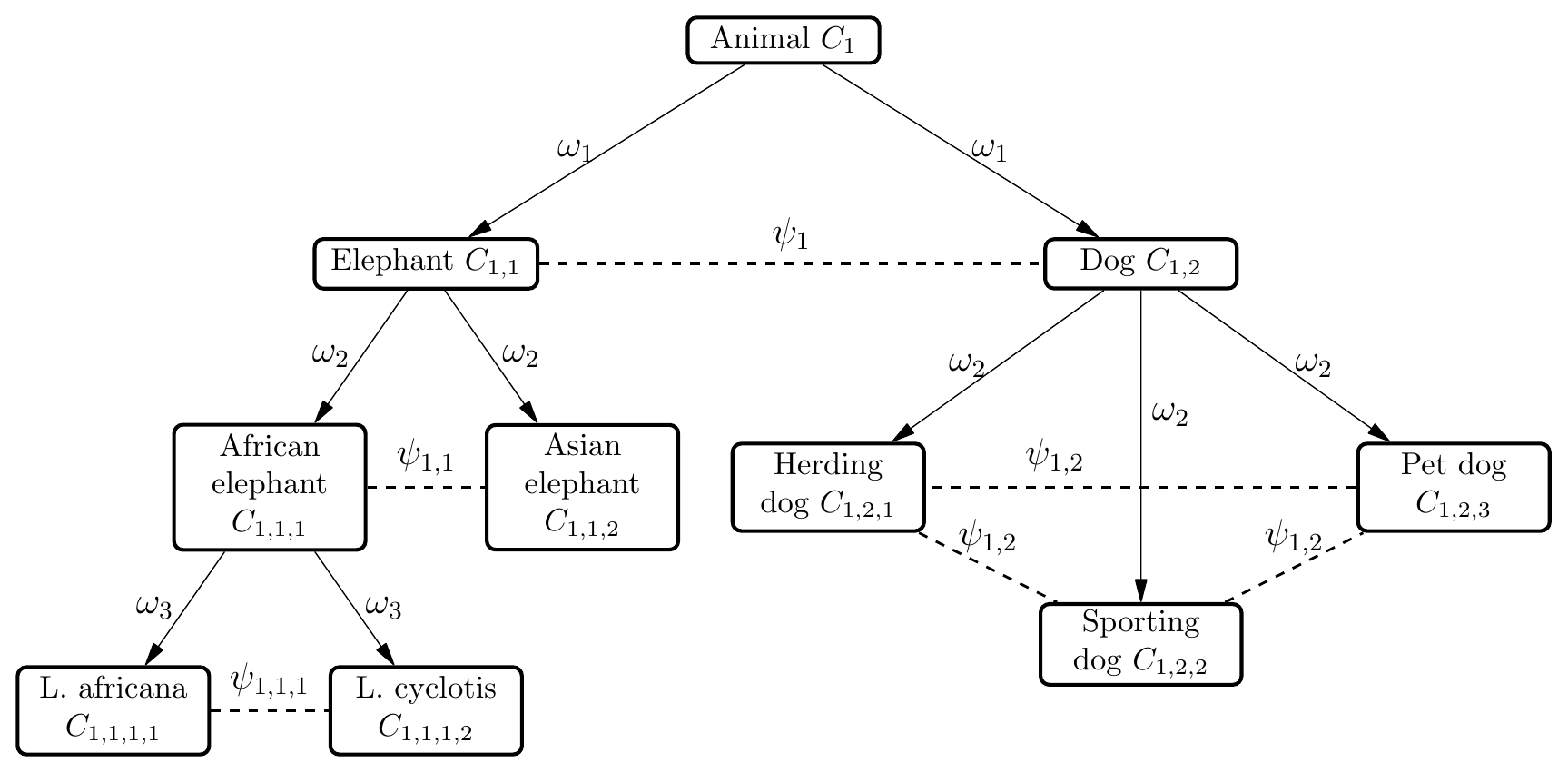} %
	\caption{An example of a hierarchical tree of 4 layers.}
\end{figure}
For the example shown in Figure 1, there are $k=4$ layers. Here, $C_1$ denotes the root node. It has two children $C_{1,1}$ and  $C_{1,2}$ with $N_{1}=2$. In addition, $C_{1,1}$ has two children  $C_{1,1,1}$ and  $C_{1,1,2}$ with $N_{1,1}=2$. Similarly, $C_{1,2}$ has three children $C_{1,2,1}$ , $C_{1,2,2}$ and $C_{1,2,3}$ with $N_{1,2}=3$. Finally, $C_{1,1,1}$ has two children $C_{1,1,1,1}$ and $C_{1,1,1,2}$ with $N_{1,1,1}=2$.

%The classical approach of hierarchical label embedding \citep{Cai:2004, Tsochantaridis:2005} works as follows.
Let $\mathcal{C}_H=\mathcal{C}\setminus\{C_1\}$ with cardinality  $q=|\mathcal{C}|-1$.
Sort the nodes  in $\mathcal{C}_H$ by layers from top to bottom,  and the nodes at the same layer from left to right. For the example in Figure 1,   nodes are ordered as $C_{1,1},C_{1,2}; C_{1,1,1}, C_{1,1,2}$, $C_{1,2,1}, C_{1,2,2}, C_{1,2,3}; C_{1,1,1,1}$, $C_{1,1,1,2}$.
Rename the ordered nodes  as $C_{(1)},\cdots, C_{(q)}$ and let $C_{(0)}$ denote the root node. The classical label embedding method \citep{Cai:2004, Tsochantaridis:2005} maps node $C_{(i)}$ into a $q$-dimensional binary vector
$\bm{u}(C_{(i)})=(u_1,\cdots, u_q)^{\top}$, where  $u_j=1$  if $C_{(j)}$  is an ancestor of $C_{(i)}$ or $j=i$, and 0 otherwise for $1\le j\le q$.
For the example in Figure 1, the embedded points of the classical method are
\begin{equation}\label{matrix_th}
\left(
\begin{array}{ccccccccc}\setlength{\arraycolsep}{-1pt}\tiny
C_{1, 1}&C_{1,2}&
C_{1,1,1}&C_{1, 1,2}&C_{1,2,1}&C_{1,2,2}&C_{1,2,3}&C_{1,1,1,1}&C_{1,1,1,2}\\
1 & 0 & 1 & 1 & 0 & 0 &0 &1 & 1 \\
0 & 1 & 0 & 0 & 1 & 1 &1 &0 & 0 \\
0 & 0 & 1 & 0 & 0 & 0 &0 &1 & 1 \\
0 & 0 & 0 & 1 & 0 & 0 &0 &0 & 0 \\
0 & 0 & 0 & 0 & 1 & 0 &0 &0 & 0 \\
0 & 0 & 0 & 0 & 0 & 1 &0 &0 & 0 \\
0 & 0 & 0 & 0 & 0 & 0 &1 &0 & 0 \\
0 & 0 & 0 & 0 & 0 & 0 &0 &1 & 0 \\
0 & 0 & 0 & 0 & 0 & 0 &0 &0 & 1 \\
\end{array}
\right).
\end{equation}
Denote $d_E(\cdot,\cdot)$ as the Euclidean distance. For any two nodes $C,C'$ in $\mathcal{C}_H$, let $d_{\mathcal{C}_H}(C,C')=d_E(\bm{u}(C),\bm{u}(C'))$.
It can be seen that  the distance between vectors in (\ref{matrix_th}) does not mimic the dissimilarity between the nodes well.
For example,  one can see that $d_{\mathcal{C}_H}(C_{1,1},C_{1,2})=d_{\mathcal{C}_H}(C_{1,1,1}, C_{1,1,2})=\sqrt{2}$.
Note that $C_{1,1,1}=\text{\{African elephant\}}$ and  $C_{1,1,2}=\text{\{Asian elephant\}}$ have the latest common ancestor $C_{1,1}=\text{\{Elephant\}}$. On the other hand, $C_{1,1}=\text{\{Elephant\}}$ and $C_{1,2}=\text{\{Dog\}}$ have the latest common ancestor $C_1=\text{\{Animal\}}$.
Since  $C_{1,1}$    is a child of $C_1$, it is more reasonable to require $d_{\mathcal{C}_H}(C_{1,1},C_{1,2})>d_{\mathcal{C}_H}(C_{1,1,1}, C_{1,1,2})$.
To proceed, we define the concept of the layer of the latest common ancestor (LLCA) as follows.
\begin{definition}  (LLCA)
For  any two nodes $C_{i_1,i_2,\cdots,i_m}$ at the $m$-th layer and $C_{j_1,j_2,\cdots,j_l}$ at the $l$-th layer, where   $i_1=j_1\equiv1$ and $1\leq m,l\leq k$,
define
$I_{i_1,i_2,\cdots,i_m;j_1,j_2,\cdots,j_l}$ as the layer at which the  latest common ancestor (LLCA) of nodes  $C_{i_1,i_2,\cdots,i_m}$ and $C_{j_1,j_2,\cdots,j_l}$ locates, that is,
$
I_{i_1,i_2,\cdots,i_m;j_1,j_2,\cdots,j_l}=
\max\Big\{t:  (i_1,i_2,\cdots,i_t)=(j_1,j_2,\cdots,j_t), 1\le t\le\min\{m,l\}\Big\}.
$
\end{definition}
Motivated from the phylogenetic tree, an ideal dissimilarity measurement denoted as $s_{\mathcal{C}}(\cdot,\cdot)$ between nodes should satisfy the following H.S. properties.
\begin{definition}(H.S. properties)\label{HS}
 \begin{itemize}

 	\item[(H.S.1)] (Hierarchical property) 	
      For two pairs  of classes   $\{C_{i_1,i_2,\cdots,i_m}$, $C_{i'_1,i'_2,\cdots,i'_{\tilde{m}}}\}$ and              $\{C_{j_1,j_2,\cdots,j_l}$, $C_{j'_1,j'_2,\cdots,j'_{\tilde{l}}}\}$,  if $I_{i_1,i_2,\cdots,i_m;i'_1,i'_2,\cdots,i'_{\tilde{m}}}<I_{j_1,j_2,\cdots,j_l;j'_1,j'_2,\cdots,j'_{\tilde{l}}}$, then  $s_{\mathcal{C}}(C_{i_1,i_2,\cdots,i_m},C_{i'_1,i'_2,\cdots,i'_{\tilde{m}}})>s_{\mathcal{C}}(C_{j_1,j_2,\cdots,j_l},C_{j'_1,j'_2,\cdots,j'_{\tilde{l}}}).$

	\item[\ \ \ (H.S.2)] (Symmetric property)
      For two pairs of classes  $\{C_{i_1,i_2,\cdots,i_m}$, $C_{i'_1,i'_2,\cdots,i'_{\tilde{m}}}\}$ and $\{C_{i_1,i_2,\cdots,i_m}$, $C_{j'_1,j'_2,\cdots,j'_{\tilde{m}}}\}$, if $I_{i_1,i_2,\cdots,i_m;i'_1,i'_2,\cdots,i'_{\tilde{m}}}=I_{i_1,i_2,\cdots,i_m;j'_1,j'_2,\cdots,j'_{\tilde{m}}}$, then $s_{\mathcal{C}}(C_{i_1,i_2,\cdots,i_m},C_{i'_1,i'_2,\cdots,i'_{\tilde{m}}})=s_{\mathcal{C}}(C_{i_1,i_2,\cdots,i_m},C_{j'_1,j'_2,\cdots,j'_{\tilde{m}}}).$
\end{itemize}
\end{definition}

The  property (H.S.1) means that if the LLCA for the   pair $C_{i_1,i_2,\cdots,i_m}$ and $C_{i'_1,i'_2,\cdots,i'_{\tilde{m}}}$ is smaller   than that of
the   pair $C_{j_1,j_2,\cdots,j_l}$ and  $C_{j'_1,j'_2,\cdots,j'_{\tilde{l}}}$, then the dissimilarity between the first pair is larger than that of the second pair.
This is similar to a phylogenetic tree.
For our example in Figure 1, $C_{1,1}$=\{Elephant\} and $C_{1,2}$=\{Dog\} have the latest common ancestor $C_{1}$ =\{Animal\}, while $C_{1,1,1}$=\{African elephant\} and $C_{1,1,2}$=\{Asian elephant\} have the latest common ancestor $C_{1,1}$=\{Elephant\}. Thus, we require  $s_{\mathcal{C}}(C_{1,1},C_{1,2})>s_{\mathcal{C}}(C_{1,1,1}, C_{1,1,2})$.
The property (H.S.2) means that for a node $C_{i_1,i_2,\cdots,i_m}$ at the $m$-th layer and other two nodes $C_{i'_1,i'_2,\cdots,i'_{\tilde{m}}}$ and
$C_{j'_1,j'_2,\cdots,j'_{\tilde{m}}}$ which are both located at the $\tilde{m}$-th layer, if the LLCA of $\{C_{i_1,i_2,\cdots,i_m}, C_{i'_1,i'_2,\cdots,i'_{\tilde{m}}}\}$ is the same as that of $\{C_{i_1,i_2,\cdots,i_m}, C_{j'_1,j'_2,\cdots,j'_{\tilde{m}}}\}$,
then the dissimilarity between the first pair  is the same as  that of the  second pair.  This property guarantees the symmetry of the dissimilarities between a node and other two nodes that locate at the same layer. For our example in Figure 1, for nodes  $C_{1,1,1,1}=$\{Loxodonta africana\}, $C_{1,2,1}=$\{Herding dog\} and $C_{1,2,2}=$\{Sporting dog\},    the   latest common ancestor for $\{C_{1,1,1,1}, C_{1,2,1}\}$ is  $C_1=$\{Animal\}, and is the same for  $\{C_{1,1,1,1}, C_{1,2,2}\}$.  Thus, it is reasonable to require
$s_{\mathcal{C}}(C_{1,1,1,1},C_{1,2,1})=s_{\mathcal{C}}(C_{1,1,1,1},C_{1,2,2})$.
In the following subsection, we define dissimilarities such that H.S. properties hold.

\subsection{Dissimilarity satisfying {\bf H.S.} properties}

A simple way to define dissimilarities between the nodes in $\mathcal{C}$ such that H.S. properties hold is desired.
To this end, we first construct a graph and  assign  two basic dissimilarities by the following two steps:

\begin{itemize}
	\item[Step 1.] (Between a parent and a child)
Add an edge between  any non-leaf node $C_{j_1,j_2,\cdots,j_{m-1}}$ at the  $(m-1)$-th layer ($2\le m\le k$) and any of its  children $C_{j_1,j_2,\cdots,j_m},j_m=1,\cdots, N_{j_1,j_2,\cdots,j_{m-1}}$. Define the dissimilarity $s_{\mathcal{C}}(C_{j_1,j_2,\cdots,j_{m-1}},C_{j_1,j_2,\cdots,j_m})= \omega_{m-1},$
where $\omega_{m-1}$ is a constant that will be specified later. Note that $\omega_{m-1}$ only depends on the layer that the node locates at, regardless of the node itself.

\item[Step 2.] (Between two siblings)
% All nodes in the second layer are siblings sharing the parent $C_1$. We add edges between any pair of them and assign the  dissimilarities as $\psi_1$.
 For $2\leq m \leq k$,  add an edge between any pair of siblings in \{$C_{j_1,j_2,\cdots,j_m},j_m=1,\cdots,N_{j_1,j_2,\cdots,j_{m-1}}$\} with the same parent $C_{j_1,j_2,\cdots,j_{m-1}}$, and assign the  dissimilarity between them as $\psi_{j_1,j_2,\cdots,j_{m-1}}$, which is a constant  depending on $(j_1,j_2,\cdots,j_{m-1})$  and will be specified later. That is,
$s_{\mathcal{C}}(C_{j_1,j_2,\cdots,j_{m-1},j'_m},C_{j_1,j_2,\cdots,j_{m-1},j^{''}_m})= \psi_{j_1,j_2,\cdots,j_{m-1}},$
for any $1\le j'_m\ne j_m^{''}\le N_{j_1,j_2,\cdots,j_{m-1}}, 2\leq m \leq k$.

\end{itemize}

Now we have a graph  $\mathcal{G}$ with  the nodes set $\mathcal{C}$ and the edges set $\mathcal{E}=$\{edges between a parent and any node of its children\}$\cup$\{edges between two  siblings\}.   An  example is shown  in Figure 1. %We see that $\omega_1$ is   the dissimilarity between the root node $C_1$ and any of its children $C_{1,j_2},j_2=1,\cdots,N_1$, and $\omega_2$ denotes the dissimilarity between any given node at the second layer and any of  its children at the third layer. The dissimilarity between $C_{1,1}$ and $C_{1,2}$ is $\psi_1$. For the two groups of siblings at the third layer, we have $s_{\mathcal{C}}(C_{1,1,1},C_{1,1,2})= \psi_{1,1},$ and $s_{\mathcal{C}}(C_{1,2,1},C_{1,2,2})= s_{\mathcal{C}}(C_{1,2,1},C_{1,2,3})=s_{\mathcal{C}}(C_{1,2,2},C_{1,2,3})= \psi_{1,2}.$
Then we can define the dissimilarity between any two different nodes, $C_{i_1,i_2,\cdots,i_m}$ at the $m$-th layer and $C_{j_1,j_2,\cdots,j_l}$ at the $l$-th layer, where $1\leq m,l\leq k$ and $i_1=j_1\equiv1$. Without any loss of generality, assume $m \leq l$.
Denote
$\text{Path}_{\text{min}}(C_{i_1,i_2,\cdots,i_m},C_{j_1,j_2,\cdots,j_l})$ as the path with the minimum number of connected edges  between these two nodes on the graph $\mathcal{G}$, and $\tilde{t}=I_{i_1,i_2,\cdots,i_m;j_1,j_2,\cdots,j_l}$. Specifically, there are two cases as follows:

\begin{itemize}
	\item[($i$)] If $\tilde{t}=m$, then $m<l$ and  $C_{i_1,i_2,\cdots,i_m}$ is the ancestor of $C_{j_1,j_2,\cdots,j_l}$. Then we have the following  $\text{Path}_{\text{min}}(C_{i_1,i_2,\cdots,i_m},C_{j_1,j_2,\cdots,j_l})$,
\begin{equation}\label{path_1}
C_{i_1,i_2,\cdots,i_{m}}\rightarrow C_{i_1,i_2,\cdots,i_m,j_{m+1}}\rightarrow\cdots\rightarrow C_{i_1,i_2,\cdots,i_m,j_{m+1},\cdots,j_{l}}.
\end{equation}

%\item[(ii)] If $\tilde{t}=m-1$ and $m=l$, then the two nodes are siblings. The edge between siblings has already been constructed, that is,
% $C_{i_1,i_2,\cdots,i_{m}}\rightarrow C_{j_1,j_2,\cdots,j_{l}}$.

\item[($ii$)] Otherwise, by the definition of $\tilde t$, it holds that  $(i_1,i_2,\cdots, i_{\tilde t})=(j_1,j_2,\cdots, j_{\tilde t})$. Then we have  $C_{i_1,i_2,\cdots,i_{\tilde t+1}}$ and $C_{j_1,j_2,\cdots,j_{\tilde t+1}}$ being siblings. Thus, $\text{Path}_{\text{min}}(C_{i_1,i_2,\cdots,i_m},C_{j_1,j_2,\cdots,j_l})$ is the combinations  of $\text{Path}_{\text{min}}(C_{i_1,i_2,\cdots,i_m},C_{i_1,i_2,\cdots,i_{\tilde t+1}})$, the edge between $C_{i_1,i_2,\cdots,i_{\tilde t+1}}$ and $C_{j_1,j_2,\cdots,j_{\tilde t+1}}$, and
$\text{Path}_{\text{min}}(C_{j_1,j_2,\cdots,j_{\tilde t+1}}, C_{j_1,j_2,\cdots,j_l})$,
     that is,
\begin{equation}\label{path_3}
C_{i_1,i_2,\cdots,i_{m}}\rightarrow \cdots\rightarrow C_{i_1,i_2,\cdots,i_{\tilde{t}+1}}\rightarrow C_{j_1,j_2,\cdots,j_{\tilde{t}+1}}\rightarrow \cdots \rightarrow C_{j_1,j_2,\cdots,j_l}.
\end{equation}
\end{itemize}

Based on $\text{Path}_{\text{min}}(C_{i_1,i_2,\cdots,i_m},C_{j_1,j_2,\cdots,j_l})$, we can define the dissimilarity between two nodes as follows.
\begin{definition}
{(Dissimilarity between two nodes)} Rename the nodes  along $\text{Path}_{\text{min}}(C_{i_1,i_2,\cdots,i_m},\\C_{j_1,j_2,\cdots,j_l})$ as $\nu_i,1\le i\le r$, where $r$ is the total number of nodes in the path, then define the dissimilarity
$
s_{\mathcal{C}}(C_{i_1,i_2,\cdots,i_m},C_{j_1,j_2,\cdots,j_l})=\left(\sum_{i=1}^{r-1} s^2_{\mathcal{C}}(\nu_i,\nu_{i+1})\right)^{1/2}.
$
\end{definition}

Combining with the definition in Steps 1 and 2, we get the explicit form of the dissimilarity between any two nodes as shown in Proposition \ref{def1}.
\begin{proposition}\label{def1}
The dissimilarity between nodes $C_{i_1,\cdots, i_m}$ and $C_{j_1,\cdots, j_l}$ is defined  as
\begin{equation}\label{def_diss}
\begin{split}
s_{\mathcal{C}}(C_{i_1,\cdots, i_m}, C_{j_1,\cdots, j_l})%=Dis(\text{Path}_{\text{min}}(C_{1,i_2,\cdots,i_m},C_{1,j_2,\cdots,j_l}))\\
=
\begin{cases}
\Bigg(\sum\limits_{i=\tilde{t}}^{\max\{m,l\}-1}\omega^2_i\Bigg)^{1/2},&\tilde{t}=\min\{m,l\},\\
\Bigg(\psi^2_{i_1,\cdots,i_{\tilde{t}}}+\sum\limits_{i=1}^{m-1}\omega^2_i+\sum\limits_{i=1}^{l-1}\omega^2_i-2\sum\limits_{i=1}^{\tilde{t}}\omega^2_{i}\Bigg)^{1/2},&\tilde{t}<\min\{m,l\},
\end{cases}
\end{split}
\end{equation}
where $\tilde{t}=I_{i_1,\cdots,i_m;j_1,\cdots,j_l}$ and $j_1=i_1\equiv1$.
\end{proposition}

Take the example in Figure 1 for an  illustration. The path with the minimum step between nodes $C_{1,1,1,1}$ and $C_{1,2,1}$ is
$
C_{1,1,1,1}\rightarrow C_{1,1,1}\rightarrow C_{1,1}\rightarrow C_{1,2}\rightarrow C_{1,2,1}.$
Thus, the dissimilarity between these two nodes is
$s_{\mathcal{C}}(C_{1,1,1,1}, C_{1,2,1})=\left(\psi^2_{1}+2\omega^2_2+\omega^2_3\right)^{1/2}.$
To ensure that the dissimilarity defined in (\ref{def_diss})  satisfies H.S.  properties, we need the following assumption.
\begin{assumption}\label{psi}
	Given $\omega_1>0$ and a constant $\delta>1$, assume that   $\omega_{m}=\omega_{m-1}/\delta$ for   $2 \leq m \leq k$ and that
	\begin{equation}\label{psi_2}
	\omega_{m-1}<\psi_{j_1,j_2,\cdots,j_{m-1}}\leq 2\omega_{m-1}, \quad\quad 2\le m\le k.
	\end{equation}
\end{assumption}
Assumption $\omega_{m}=\omega_{m-1}/\delta$ means that the dissimilarity between a parent  and any node of its children decreases when the layer increases along the tree, which is reasonable.
The left part of (\ref{psi_2}) means that the dissimilarity between a node and its parent is not larger than that between it and its sibling. The right part of (\ref{psi_2}) is in a similar spirit as that of the triangle inequality.
In Section 3.2, we give a specific form of $\psi_{j_1,j_2,\cdots,j_{m-1}}$ satisfying Assumption \ref{psi}. By Assumption 1, the following Theorem 1 shows that H.S. properties hold.

\bet\label{Th1-DHC}
Under  Assumption \ref{psi}  with $\delta^2 \geq 2\sqrt{2}+2$, the dissimilarity defined in (\ref{def_diss}) satisfies H.S. properties.
\eet

\section{Exact label embedding}
In this section, we consider  the exact label embedding in  hierarchical classification, by establishing an isometry (i.e. the Euclidean distance between  embedded points is exactly equal to the dissimilarity between nodes on the tree).  We first consider the case of $\tilde q$-class multicategory classification, where the  label embedding approach has been considered in literature \citep{Lange:2008, Zhang:2014}. We give a different way to construct points for multicategory classification in Section 3.1, and then extend to hierarchical classification in Section 3.2.

\subsection{Label embedding in  multicategory classification}

In this subsection, we give the procedure to construct  $\tilde{q}$ $(\tilde{q}\geq 2)$ points with equal pairwise distances in the $\mR^{\tilde{q}-1}$ Euclidean space, which  is similar in spirit to the methods in \citet{Lange:2008}, \citet{ Wu:2010}, \citet{Wu:2012} and  \citet{Zhang:2014} for $\tilde{q}$-class multicategory classification problems.
%For any vector $\bm{\xi}\in \mR^{\tilde{q}-1}$, recall $\bm{\xi}^{(m)}$ is the subvector consisting of the first $m$ coordinates of $\bm{\xi}$. Let  $\{e_i\}_{i=1}^{\tilde{q}-1}$ be the coordinate bases of $\mR^{\tilde q-1}$, i.e. $e_i=(0,\cdots, 0, 1,0,\cdots,0)^{\top}$ with the $i$-th coordinate being 1.
The $\tilde q$ points $\{\bm{\xi}_i\}_{i=1}^{\tilde{q}}$  in $\mR^{\tilde{q}-1}$ with an equal pairwise distance can be constructed  by Algorithm 1, which indeed formulate a simplex. Recall that $\bm{u}^{(m)}$ denotes the subvector consisting of the first $m$ coordinates of $\bm{u}$.

\begin{algorithm}[t]
\caption{\bf: Label embedding in  multicategory classification}
\begin{itemize}
  \item[ 1.]\textit{Initialization:}   Given a constant $c>0$,  set $\xi_1^{(1)}=c/2, \xi_2^{(1)}=-c/2$.

  \item[ 2.]\textit{Iteration:}  For $m=2,\cdots, \tilde{q}-1$, repeat the  following steps (1) and (2).
              \begin{itemize}
              \item [(1)]  Set $\bm{\xi}_{i}^{(m)}=((\bm{\xi}_{i}^{(m-1)})^{\top},0)^{\top} \in \mR^m, i=1,\cdots,m$;
              \item [(2)] $\bm{\xi}_{m+1}^{(m)}=m^{-1}\sum_{i=1}^m \bm{\xi}_i^{(m)}+a_m \bm{e}_{m}^{(m)}$, where   $a_m=\sqrt{c^2-d_{m-1}^2}$  with $d_{m-1}=\|m^{-1}\sum_{i=1}^m \bm{\xi}_i^{(m-1)}-\bm{\xi}_m^{(m-1)}\|$, and $\bm{e}_m\in\mathbb{R}^m$ with the $m$-th coordinate being 1 and others being 0.
              \end{itemize}
  \item[ 3.]\textit{Centralization:}  Let  $\bm{\xi}_{i}\leftarrow \bm{\xi}_{i}-(\tilde{q})^{-1}\sum_{j=1}^{\tilde{q}} \bm{\xi}_{j},i=1,\cdots,\tilde{q}$.
                %The points $\bm{\xi}_i$'s are of equal pairwise distance $c$.

  \item[4.]\textit{Scaling:}   $\tilde{\bm{\xi}}_i\leftarrow T T_{\tilde{q}}^{-1} \bm{\xi}_i$ for $1\le i\le \tilde{q}$, where $T_{\tilde q}$ is given in Proposition \ref{prop3}. %This is an optional step. After this step, the   constructed  points have a fixed $l_2$ norm $T$ and an equal pairwise distance $c_{\tilde q}$ defined in Proposition \ref{prop3}.

\end{itemize}
\end{algorithm}

\bep\label{prop3} The following conclusions hold.
\begin{itemize}
  \item[(1)] For $\{\bm{\xi}_i\}_{i=1}^{\tilde{q}}$ constructed in Steps 1-3 of Algorithm 1,
        \begin{itemize}
        \item[(i)] $\|\bm{\xi}_i-\bm{\xi}_j\|=c$ for $1\leq i \neq j \le \tilde{q}$;
        \item[(ii)] $\|\bm{\xi}_{i}\| = T_{\tilde{q}}$ for $1\leq i \leq \tilde{q}$, where $T_{\tilde{q}} =c[(\tilde{q}-1)/2\tilde{q}]^{1/2}$;
        \item[(iii)] the angles  $\angle(\bm{\xi}_{i}, \bm{\xi}_{j})$ are all equal for $1\leq i \neq j \le \tilde{q}$ with $\cos{\angle(\bm{\xi}_{i},\bm{\xi}_{j})}=-1/(\tilde{q}-1)$.
        \end{itemize}

  \item[(2)]  For $\{\tilde {\bm{\xi}}_i\}_{i=1}^{\tilde{q}}$ constructed in Steps 1-4 of Algorithm 1,
  \begin{itemize}
  	\item[(i)]   $\|\tilde{\bm{\xi}}_{i}\|=T$  for $1\leq i \leq \tilde{q}$;
  	\item[(ii)] $\|\tilde{\bm{\xi}}_{i}-\tilde{\bm{\xi}}_{j}\| = c_{\tilde{q}}$ for $1\leq i\ne j \leq \tilde{q}$, where $c_{\tilde{q}}=T[2\tilde{q}/(\tilde{q}-1)]^{1/2}$.
   \end{itemize}
\end{itemize}
\eep

Proposition \ref{prop3} shows that $\{\bm{\xi}_i\}_{i=1}^{\tilde q}$ constructed in Steps 1--3  have an equal pairwise distance $c$.  Thus, if constructing  points of an equal pairwise distance $c$ is the goal, implementing Steps 1--3 is sufficient.  In some cases, it is desirable to require further the constructed points having the same   norm $T$. To this end, in Step 4, we scale the points $\{\bm{\xi}_i\}_{i=1}^{\tilde{q}}$. After Step 4, the points $\{\tilde{\bm{\xi}}_i\}_{i=1}^{\tilde{q}}$ have the equal pairwise distance $c_{\tilde q}$ and the same $l_2$ norm $T$.

\begin{remark}
Points constructed in Algorithm 1 are in the  space spanned by  $\{\bm{e}_j\in \mR^{\tilde q-1}, 1\le j\le \tilde q\}$, where $\bm{e}_j$'s are the coordinate bases of $\mR^{\tilde q-1}$. For integers $a$ and $b$ with  $b\ge a+\tilde q-1$ and $a\ge 0$, Algorithm 1 can be directly extended to construct points in the subspace spanned by the coordinate bases $\{ \bm{e}_j\in \mR^{b},   a+1\le j\le a+\tilde q-1\}$,
 by extending the vector $\tilde {\bm{\xi}}_i$ to $(\bm{0}_{a}^{\top},\tilde{\bm{\xi}}_i^{\top}, \bm{0}_{b-a-\tilde q+1}^{\top})^{\top}$, $1\le i \le \tilde q$.
\end{remark}

For the example in Figure 1, if we ignore the tree structure and consider multicategory classification on leaf nodes, the points constructed by Algorithm 1 given $T=1$ are
\begin{equation}\label{matrix_tm}
\left(
\begin{array}{cccccc}
C_{1,1,1,1}&C_{1,1,1,2}&C_{1, 1,2}&C_{1,2,1}&C_{1,2,2}&C_{1,2,3}\\
-\sqrt{15}/5 & \sqrt{15}/5 & 0 & 0&0&0 \\
-\sqrt{5}/5 & -\sqrt{5}/5 & 2\sqrt{5}/5 & 0 & 0 &0\\
-\sqrt{10}/10 & -\sqrt{10}/10 & -\sqrt{10}/10& 3\sqrt{10}/10 &0&0\\
-\sqrt{6}/10 & -\sqrt{6}/10 & -\sqrt{6}/10  & -\sqrt{6}/10 & 2\sqrt{6}/5 & 0\\
-1/5 & -1/5 & -1/5  & -1/5 & -1/5 & 1\\
\end{array}
\right).
\end{equation}
The points in (\ref{matrix_tm}) incorporate no hierarchy, and the distance is $2\sqrt{15}/5$ for all pairs of leaf nodes. As stated in (H.S.1), it is more reasonable to require $d_{\mathcal{C}_H}(C_{1,1,1}, C_{1,1,2})<d_{\mathcal{C}_H}(C_{1,1,2},C_{1,2,1})$.
%since  $C_{1,1}=\text{\{Elephant\}}$, the latest common ancestor of the pair $(C_{1,1,1},C_{1,1,2})$,  is    a child of  $C_{1}=\{\text{Animal}\}$, which is  the latest common ancestor  of the pair  $(C_{1,1,2},C_{1,2,1})$.
In the following subsection, we propose an approach, which incorporates the hierarchical information  while requiring the same  dimension as (\ref{matrix_tm}).

\subsection{Label embedding in hierarchical classification}

We now extend the idea in Section 3.1 to hierarchical classification.
Note that the root is meaningless, and we ignore it.
Recall $n_{\mathrm{leaf}}$ is the number of leaf nodes on the tree and $j_1\equiv1$. For $2\leq m \leq k$,  denote the  nodes  at the $m$-th layer and their corresponding embedding  points respectively as $\mathcal{C}_{H,m}=\{C_{j_1,j_2,\dots, j_m},  j_s=1,\cdots , N_{j_1,j_2,\dots, j_{s-1}},s=2,\cdots,m\},E_{H,m}=\{\bm{\xi}_{j_1,j_2,\dots, j_m}\in \mR^{K}:  j_s=1,\cdots , N_{j_1,j_2,\dots, j_{s-1}},s=2,\cdots,m\},$
where the  dimension $K\ge n_{\mathrm{leaf}}-1$. In fact, Proposition \ref{Prop5} below shows that it is sufficient to set $K=n_{\mathrm{leaf}}-1$.
%Moreover, by our approach,  for any point in $E_{H,m}$, only the first $D_m$ coordinates are nonzero, where $D_m$ will   be specified later.

%Specifically, we start by initializing all the points $\bm{\xi}_{j_1,j_2,\dots, j_m}$'s  being a zero vector.
For $m=2$, there are  $N_{j_1}$ nodes in $\mathcal{C}_{H,2}$. Let  $D_2=N_{j_1}-1$. Then we construct points $\{\bm{\xi}_{j_1,j_2}\in \mathbb{R}^{K}, j_2=1,\cdots, N_{j_1}\}$, where the subvectors $\{\bm{\xi}_{j_1,j_2}^{(D_2)}, j_2=1,\cdots, N_{j_1}\}$ of dimension $D_2$ are constructed by Algorithm 1 with a given norm $T^{(1)}$, and the coordinates of $\bm{\xi}_{j_1j_2}$ with indices  larger than $D_2$   are set zero.
For $m=3,\cdots,k$, $E_{H,m}$   is  constructed by Algorithm 2. We see that the $i$-th ($i> D_k$) coordinate of any point $\bm{\xi}_{j_1,j_2,\cdots,j_m}$ is zero. Furthermore, Proposition \ref{Prop5} shows that $D_k=n_{\text{leaf}}-1$. Thus, it is sufficient to set $K=n_{\text{leaf}}-1$.

\begin{algorithm}[h]
\caption{\bf: Label embedding in hierarchical classification}
For $m=3,\cdots, k$, repeat the following Steps 1-3:
\begin{itemize}
  \item[1.] Sort all  non-leaf nodes in $\mathcal{C}_{H,m-1}$ from left to right and rename them as $C_1^{(m-1)},\cdots, C_{n_{m-1}}^{(m-1)}$,  where  $n_{m-1}$ is the number of non-leaf nodes at the $(m-1)$-th layer. Then for any $1\le i\le n_{m-1}$, there exists some index $(j_2',\cdots, j_{m-1}')$ such that
  $C_i^{(m-1)}=C_{j_1',j_2',\cdots, j_{m-1}'}$ with $j_1'\equiv1$. For each non-leaf node $C_i^{(m-1)}=C_{j_1',j_2',\cdots, j_{m-1}'}$, it has children  $\mathrm{Chi}(C_i^{(m-1)})=\{C_{j_1',j_2',\cdots, j_{m-1}', j_{m}},  j_{m}=1,\cdots, N_{j_1',j_2',\cdots, j_{m-1}'}\}$ at the $m$-th layer with  $N_{j_1',j_2',\cdots, j_{m-1}'}\geq 2$ according to our assumption that each parent node has at least two children. Let   $d_{m,i}=N_{j_1',j_2',\cdots, j_{m-1}'}-1$, $i=1,\cdots, n_{m-1}$ and $T^{(m-1)}=T^{(m-2)}/\delta$ with $\delta$ being  the  constant  defined in Assumption \ref{psi}.

  \item [2.]  For any $C_i^{(m-1)}$ ($1\leq i\leq n_{m-1}$) and its children $\mathrm{Chi}(C_i^{(m-1)})$, we construct $N_{j_1',j_2',\cdots, j_{m-1}'}$ points  denoted  as $\{\bm{\eta}_{j_1',j_2',\cdots, j'_{m-1},j_{m}}, j_{m}=1,\cdots, N_{j_1',j_2',\cdots, j_{m-1}'}\}$ based on Algorithm 1 and Remark 1 in Section 3.1 with the given norm $T^{(m-1)}$ in the subspace
  \begin{equation}\label{space}
  \mathrm{span}\left\{\bm{e}_j \in \mathbb{R}^{K}:D_{m-1}+1+\sum\limits_{s=0}^{i-1} d_{m,s} \le j \le  D_{m-1}+\sum\limits_{s=0}^i d_{m,s}\right\}
  \end{equation}
  where  $d_{m,0}=0$ and $\bm{e}_j$'s are the coordinate bases in $\mathbb{R}^{K}$. Then let
  \begin{equation}\label{inherit}
  \bm{\xi}_{j_1',j_2',\cdots,j'_{m-1}, j_{m}}=\bm{\xi}_{j_1',j_2',\cdots, j_{m-1}'}+\bm{\eta}_{j_1',j_2',\cdots, j'_{m-1}, j_{m}},\quad j_{m}=1,\cdots, N_{j_1',j_2',\cdots, j_{m-1}'}.
  \end{equation}

  \item[3.] Repeat Step 2 for all $n_{m-1}$ non-leaf nodes in $\mathcal{C}_{H,m-1}$ and set
  $
  D_{m}=D_{m-1}+\sum_{i=1}^{n_{m-1}} d_{m,i}.
  $
\end{itemize}
\end{algorithm}

\bep\label{Prop5} It holds that
$D_k=n_{\mathrm{leaf}}-1$ and for any  $i>D_k$,
the $i$-th coordinate of any  point $\bm{\xi}_{j_1,j_2,\cdots, j_m}\in  \bigcup\limits_{l=2}^k E_{H,l}$ is zero. Thus, the dimension of  the embedded space can be set as $K=D_k=n_{\mathrm{leaf}}-1$.
\eep

%From the proof of Proposition \ref{Prop5}, we see  $D_{m}=\sum\limits_{l=2}^m\left(|E_{H,l}|-n_{l-1}\right)$.
Note that for a different $i$, the subspaces (\ref{space}) are orthogonal. The coordinates of $\bm{\xi}_{j_1',j_2',\cdots, j_{m-1}'}$ with an index larger than $D_{m-1}$ are all zero by the proof of  Proposition \ref{Prop5},  while the first $D_{m-1}$ coordinates of $\bm{\eta}_{j_1',j_2',\cdots, j'_{m-1}, j_{m}}$ are all zero. Thus, (\ref{inherit}) in Step 3 indicates that the constructed points inherit the coordinates from its parent node, and then incorporate the  hierarchical information.
Recall $\mathcal{C}_H=\mathcal{C}/\ \{C_1\}=\bigcup\limits_{m=2}^k\mathcal{C}_{H,m}$. Let  $E_H=\bigcup\limits_{m=2}^kE_{H,m}$.
Define the map $
F_H: \mathcal{C}_H\rightarrow E_H$ satisfying
$
F_H(C_{j_1,j_2,\cdots, j_m})=\bm{\xi}_{j_1,j_2,\cdots, j_m}.
$
Then the following theorem shows that $F_H$ is an isometry.

\bet\label{Th2-isometry}
Assume that $\delta^2\geq2\sqrt{2}+2$. Let
$
\psi_{j_1,j_2,\cdots,j_{m-1}}=\omega_{m-1}[2N_{j_1,\cdots,j_{m-1}}/(N_{j_1,\cdots,j_{m-1}}-1)]^{1/2},2 \leq m \leq k,
$
where $j_1\equiv1$. Then $\psi_{j_1,j_2,\cdots,j_{m-1}}$ satisfies Assumption \ref{psi}. In addition, for any two nodes $C_{i_1,i_2,\cdots, i_{m}}$ and $C_{j_1,j_2,\cdots, j_{l}}$, it holds that
$$s_\mathcal{C}(C_{i_1,i_2,\cdots, i_{m}}, C_{j_1,j_2,\cdots, j_{l}})=\frac{\omega_1}{T^{(1)}} d_E(F_H(C_{i_1,i_2,\cdots, i_{m}}),F_H(C_{j_1,j_2,\cdots, j_{l}})).
$$
Specifically, setting $T^{(1)}=\omega_1$, we have that $F_H$ is an isometry from $(\mathcal{C}_H,s_{\mathcal{C}})$ to $(E_H, d_E)$.
\eet

\begin{remark}
%Note that $\omega_1$ is the dissimilarity between the root node and its children in the second layer and  $\omega_m=\omega_{m-1}/\delta$.
Without loss of generality, we set $\omega_1=1$ and consequently $T^{(1)}=1$ to keep the isometry property.   In fact, classification results are invariant for any  $T^{(1)}>0$ according to the results in  Sections 4 and 5.  Moreover, as long as $\delta^2\geq2\sqrt{2}+2$, preliminary experiments show that the effect of $\delta$ is limited.  In this paper, we set $\delta=\sqrt{5}$. In addition,  $\delta$ is a constant independent of $m$, and it can be extended to allow $\delta$ depending on $m$, i.e.  $\omega_m=\omega_{m-1}/\delta_m$.
\end{remark}

Combining Theorems \ref{Th1-DHC} and \ref{Th2-isometry}, we can show that the embedded points also satisfy H.S. properties as stated in the following proposition.

\bep  For $\delta^2\geq2\sqrt{2}+2$, it holds that
\begin{itemize}
	\item[(1)] For any two pairs of  points $\{\bm{\xi}_{i_1,\cdots,i_m},\bm{\xi}_{i'_1,\cdots,i'_{\tilde{m}}}\}$ and $\{\bm{\xi}_{j_1,\cdots,j_l},\bm{\xi}_{j'_1,\cdots,j'_{\tilde{l}}}\}$, if $I_{i_1,\cdots,i_m;i'_1, \cdots,i'_{\tilde{m}}}<I_{j_1,\cdots,j_l;j'_1,\cdots,j'_{\tilde{l}}}$, then
	$
	d_{E}(\bm{\xi}_{i_1,\cdots,i_m},\bm{\xi}_{i'_1,\cdots,i'_{\tilde{m}}})>d_E(\bm{\xi}_{j_1,\cdots,j_l},\bm{\xi}_{j'_1,\cdots,j'_{\tilde{l}}}).
	$
	\item[(2)] For any two pairs of   points  $\{\bm{\xi}_{i_1,\cdots,i_m},\bm{\xi}_{i'_1,\cdots,i'_{\tilde{m}}}\}$ and $\{\bm{\xi}_{i_1,\cdots,i_m},\bm{\xi}_{j'_1,\cdots,j'_{\tilde{m}}}\}$, if $I_{i_1,\cdots,i_m;i'_1, \cdots,i'_{\tilde{m}}}=I_{i_1,\cdots,i_m;j'_1,\cdots,j'_{\tilde{m}}}$, then
	$
	d_{E}(\bm{\xi}_{i_1,\cdots,i_m},\bm{\xi}_{i'_1,\cdots,i'_{\tilde{m}}})=d_E(\bm{\xi}_{i_1,\cdots,i_m},\bm{\xi}_{j'_1,\cdots,j'_{\tilde{m}}}).
	$
\end{itemize}
\eep

For the  example in Figure 1, the whole label embedding matrix is constructed as
\begin{equation}\label{matrix_h}
\left(
\begin{array}{ccccccccc}
C_{1,1}&C_{1,2}&
C_{1,1,1}&C_{1, 1,2}&C_{1,2,1}&C_{1,2,2}&C_{1,2,3}&C_{1, 1,2,1}&C_{1,1,2,2}\\
-1 & 1 & -1 & -1 & 1 & 1 &1 &-1 & -1 \\
0 & 0 & -\sqrt{5}/5 & \sqrt{5}/5 & 0 & 0 &0& -\sqrt{5}/5& -\sqrt{5}/5\\
0 & 0 & 0 & 0 & -\sqrt{15}/10 & \sqrt{15}/10&0& 0 & 0\\
0 & 0 & 0  & 0 & -\sqrt{5}/10 & -\sqrt{5}/10&\sqrt{5}/5 & 0& 0\\
0 & 0 & 0  & 0 & 0 & 0 & 0& -1/5&1/5\\
\end{array}
\right).
\end{equation}
The dimension of the embedding space is   5, much smaller than 9 required by the classical label embedding method (\ref{matrix_th}), and is  the same as that of the multicategory case in (\ref{matrix_tm}), where the hierarchical information is ignored.  The distance matrix associated with (\ref{matrix_h}) is
$$
\left(
\begin{array}{ccccccccc}
&C_{1,2}&
C_{1,1,1}&C_{1, 1,2}&C_{1,2,1}&C_{1,2,2}&C_{1,2,3}&C_{1, 1,2,1}&C_{1,1,2,2}\\
C_{1,1}&2& \sqrt{5}/5 &\sqrt{5}/5 &\sqrt{105}/5 &\sqrt{105}/5&\sqrt{105}/5 &\sqrt{6}/5 & \sqrt{6}/5\\
C_{1,2}&&\sqrt{105}/5&\sqrt{105}/5&\sqrt{5}/5&\sqrt{5}/5&\sqrt{5}/5
&\sqrt{106}/5&\sqrt{106}/5
\\
C_{1,1,1}&&&2\sqrt{5}/5&\sqrt{110}/5&\sqrt{110}/5&\sqrt{110}/5
&1/5&1/5\\
C_{1,1,2}&&&&\sqrt{110}/5&\sqrt{110}/5&\sqrt{110}/5
&\sqrt{21}/5&\sqrt{21}/5\\
C_{1,2,1}&&&&&\sqrt{15}/5&\sqrt{15}/5
&\sqrt{111}/5&\sqrt{111}/5\\
C_{1,2,2}&&&&&&\sqrt{15}/5&\sqrt{111}/5&\sqrt{111}/5\\
C_{1,2,3}&&&&&&&\sqrt{111}/5&\sqrt{111}/5\\
C_{1,1,2,1}&&&&&&&&2/5\\
\end{array}
\right).
$$
It can be seen that our embedding points incorporate the hierarchy. Recall that $C_{1,1}=\text{\{Elephant\}},C_{1,2}=\text{\{Dog\}}$, $C_{1,1,1}=\text{\{African elephant\}}, C_{1,1,2}=\text{\{Asian elephant\}}$, $C_{1,2,1}=\text{\{Herding  dog\}}$.  Then
we have $d_{\mathcal{C}_H}(C_{1,1},C_{1,2})=2$, which is larger than $d_{\mathcal{C}_H}(C_{1,1,1}, C_{1,1,2})=2\sqrt{5}/5$, and the latter is  smaller than $d_{\mathcal{C}_H}(C_{1,1,2},C_{1,2,1})=\sqrt{110}/5$.

\section{Angle-based hierarchical classification via exact label embedding}

In hierarchical classification, denote $\bm{Z}=(\bm{X},Y)\in\mathcal{X}\times \mathcal{Y}$, where $\mathcal{X}\subset \mR^p$ and $\mathcal{Y}$ is the set of paths from the root to a leaf on the tree. Specifically, $\bm{X}\in \mR^p$ is a $p$-dimensional input vector, and $Y=\{Y^{(1)},\cdots,Y^{(\mathcal{L}(Y))}\}$ is the corresponding output with $Y^{(m)}$ indicating the label at the $m$-th layer and $\mathcal{L}(Y)$ being the layer where the leaf locates, i.e. $Y^{(1)}=C_1$,  and $Y^{(m)}\in\text{Chi}(Y^{(m-1)})$ for $m=2,\cdots,\mathcal{L}(Y)$. For example, two possible paths in Figure 1 are $y=\{C_1,C_{1,1},C_{1,1,1},C_{1,1,1,1}\}$ with $\mathcal{L}(y)=4$ and $y=\{C_1,C_{1,2},C_{1,2,1}\}$ with $\mathcal{L}(y)=3$.

\subsection{Hierarchical classification  with the top-down strategy}

Different from multicategory classification, the strategies for hierarchical classification are more complicated.
%Note that an observation $x$ can be  classified by different strategies to coincide with the hierarchy which can lead to different feasible sets of estimates.
The most commonly used strategy is the top-down (TD) \citep{wang2011large}, which is adopted in this paper. It means given an instance classified to a node at the $(m-1)$-th layer, we only consider to classify it
into one of its children at the $m$-th layer.
%
%The top-down methods classify a sample by filtering it down a tree from the root node.
%For each parent node where this sample arrives, those child nodes
%whose confidence scores (produced by the classifiers)
%pass the thresholding strategies will carry it on.
%The bottom-up approach emphasizes upon the emergence of a
%learning algorithm from the lower level nodes, without an
%underlying emphasis on the type of learning to be designed forming the basis of the modelling
%process.
%

For $m=2,\cdots,\mathcal{L}(y)$, define   $\bm{\xi}_m(y)=F_H(y^{(m)})$ being   the corresponding constructed point, i.e.   $\bm{\xi}_m(y)=\bm{\xi}_{j_1,\cdots,j_m}$ if $y^{(m)}=C_{j_1,\cdots,j_m}$.
Denote the learning function as $\bm{f}(\bm{x})=(f_1(\bm{x}),\cdots, f_K(\bm{x}))^\top\in \mR^K$, where $K=n_{\text{leaf}}-1$.
%with each coordinate of $x$ being bounded by 1 which can always be achieved by centralization and some linear transformation.
Similar to Zhang and Liu (2014), we define a linear discriminant function $
g: (\bm{f}(\bm{x}),\bm{\xi}_m(y))\rightarrow \langle \bm{f}(\bm{x}),\bm{\xi}_m(y)\rangle,
$
where $\langle \cdot,\cdot\rangle$ denotes the inner product in the Euclidean space.
We denote by $\hat{y}=\mathcal{H}(\bm{f}(\bm{x})) \in \mathcal{Y}$ the predicted path for $\bm{x}$ according to the following hierarchical classification rule with given $\bm{f}(\bm{x})$.
Note that $\hat{y}^{(1)}\equiv C_{1}$.
%Suppose that $x$ has been assigned with $y^{(m-1)}$ at the $(m-1)$-th layer. We consider the following TD hierarchical classification strategy  to assign a label at the $m$-th layer.
\begin{definition}
	{(TD)} For $m\geq 2$, assume $\bm{x}$ has been assigned to the label $\hat{y}^{(m-1)}$ at the $(m-1)$-th layer. When $\hat{y}^{(m-1)}$ is not a leaf node, we
	assign $x$ to one of its children $\hat{y}^{(m)}\in \text{Chi}(\hat{y}^{(m-1)})$ at the $m$-th layer, if the corresponding embedded point $\bm{\xi}_m(\hat{y})$ has the largest inner product at the $m$-th layer, that is, for any $\tilde{y}\in \mathcal{E}_{m}(\hat{y})=\{\tilde{y} :\tilde{y}^{(m)}\ne \hat{y}^{(m)},  \tilde{y}^{(m)}\in \text{Chi}(\hat{y}^{(m-1)})\}$,
	\begin{equation}\label{d_prin}
	g(\bm{f}(\bm{x}),\bm{\xi}_m(\hat{y}))\ge g(\bm{f}(\bm{x}),\bm{\xi}_m(\tilde{y})).
	\end{equation}
\end{definition}

Note that $\bm{\xi}_m(\tilde{y})$ depends only on $\tilde{y}^{(m)}$.  It is possible that there are many $\tilde y\in \mathcal{E}_{m}(\hat{y})$ with the same label $\tilde y^{(m)}$  at the $m$-th layer. If this is the case, taking only one of them as the representative and denoting the set of  representatives as  $[\mathcal{E}_{m}(\hat{y})]$, we only need to require that (\ref{d_prin}) holds for any $\tilde{y}\in [\mathcal{E}_{m}(\hat{y})]$.
Take the example in Figure 1 for an illustration. Let $\hat{y}^{(1)}=C_1$. We consider the children  $C_{1,1}$ and $C_{1,2}$.  Then $\bm{x}$ is assigned with $\hat{y}^{(2)}=C_{1,1}$, if (\ref{d_prin}) holds for any $\tilde{y}\in \mathcal{E}_{2}(\hat{y})$ with $\mathcal{E}_{2}(\hat{y})=\{\{C_1,C_{1,2},C_{1,2,1}\},\{C_1,C_{1,2},C_{1,2,2}\},\{C_1,C_{1,2},C_{1,2,3}\}\}$. The three paths in $\mathcal{E}_{2}(\hat{y})$ have the same label $C_{1,2}$ at the second layer, then it is sufficient to take any of them as the representative, and require (\ref{d_prin}) holds, i.e. $\langle\bm{f}(\bm{x}),\bm{\xi}_{1,1}\rangle>\langle\bm{f}(\bm{x}),\bm{\xi}_{1,2}\rangle$; otherwise,  we set $\hat{y}^{(2)}=C_{1,2}$. Supposing $\hat{y}^{(2)}=C_{1,2}$, we then consider the children of $C_{1,2}$, that is, $C_{1,2,1},C_{1,2,2},C_{1,2,3}$. Computing  $\langle\bm{f}(\bm{x}),\bm{\xi}_{1,2,1}\rangle, \langle\bm{f}(\bm{x}),\bm{\xi}_{1,2,2}\rangle$ and $\langle\bm{f}(\bm{x}),\bm{\xi}_{1,2,3}\rangle$,  $\hat{y}^{(3)}$ is taken as the class that the corresponding embedded point has the largest inner product.

It is shown in Lemma 2 of the Supplementary Material that all points  $\bm{\xi}_m(\tilde{y})$ with $\tilde y\in [\mathcal{E}_m(\hat{y})]$ have the same norm. Therefore, (\ref{d_prin}) holds as long as
$ d_{E}(\bm{f}(\bm{x}),\bm{\xi}_m(\hat{y}))\le d_{E}(\bm{f}(\bm{x}),\bm{\xi}_m(\tilde{y}))$,
which shows that our classification  strategy is essentially based on the Euclidean distance.
Given the linear discriminant function $g$, define  $G_{m}(\bm{f}(\bm{x}),y,\tilde{y})=g(\bm{f}(\bm{x}),\bm{\xi}_m(y))-g(\bm{f}(\bm{x}),\bm{\xi}_m(\tilde{y})), \tilde{y}\in[\mathcal{E}_{m}(y)], m=2,\cdots, \mathcal{L}(y).$
For an instance $z=(\bm{x},y)$, we define the following {\it hierarchy margin} $M(\bm{f}(\bm{x}),y)$ associated with the strategy TD,
\begin{align*}
M(\bm{f}(\bm{x}),y)=&\min_{m=2,\cdots,\mathcal{L}(y)}\left[g(\bm{f}(\bm{x}),\bm{\xi}_m(y))-\max\limits_{\tilde{y}\in [\mathcal{E}_m(y)]}g(\bm{f}(\bm{x}),\bm{\xi}_m(\tilde{y}))\right]\\
=& \min_{m=2,\cdots, \mathcal{L}(y),\tilde y\in [\mathcal{E}_m(y)]} G_{m}(\bm{f}(\bm{x}),y,\tilde{y}).
\end{align*}

A positive margin is required for the classifier to assign the correct label along  the whole path, which is equivalent to a set of linear constraints
\begin{equation}\label{p-margin}
G_{m}(\bm{f}(\bm{x}),y,\tilde{y})\geq 0, \quad \tilde{y}\in [\mathcal{E}_m(y)], m=2,\cdots, \mathcal{L}(y).
\end{equation}

A special case of the hierarchy margin  is $k=2$  without any hierarchical structure. If $M(\bm{f}(\bm{x}),y)\geq 0$, that is, $\langle \bm{f}(\bm{x}),\bm{\xi}_2(y) \rangle\ge \langle \bm{f}(\bm{x}),\bm{\xi}_2(\tilde{y})\rangle$ for any $\tilde{y}\in [\mathcal{E}_2(y)]=\{\tilde{y}:\tilde{y}^{(2)}\neq y^{(2)} \}$, then $\bm{x}$ is correctly classified by $\bm{f}(\bm{x})$. This is exactly the same as the method of \citet{Zhang:2014} for multicategory classification. Therefore, the hierarchy margin is a natural extension of the margin in multicategory classification \citep{Zhang:2014}.

Recall $\mathcal{H}(\bm{f}(\bm{x}))\in \mathcal{Y}$ is the classification rule by $\bm{f}(\bm{x})$ with the top-down strategy.
There are several definitions of the generalization error in hierarchical classification, and details can be referred to \citet{Wang:2009,wang2011large,babbar2016learning}.
In this paper,  we use the  0-1 hierarchical loss and define $R(\bm{f})=E[I(Y\neq\mathcal{H}(\bm{f}(\bm{X})))]$ \citep{wang2011large,babbar2016learning}. One can verify that $I(Y\neq\mathcal{H}(\bm{f}(\bm{X})))=I(M(\bm{f}(\bm{X}),Y)<0)$. A classification error occurs if   (\ref{p-margin}) fails, that is, $G_{m}(\bm{f}(\bm{x}),y,\tilde{y})< 0$ for some $m$ and $\tilde{y}\in [\mathcal{E}_m(y)]$.

Let $\{\bm{z}_i: \bm{z}_i=(\bm{x}_i, y_i)\}^n_{i=1}$ be a set of $n$ labeled training samples.
The empirical generalization error is defined as $n^{-1}\sum_{i=1}^nI(M(\bm{f}(\bm{x}_i),y_i)<0)$, which is computationally infeasible because of the discontinuity. %Commonly used surrogate loss functions include the  hinge loss $\ell_{\text{hinge}}(u)=(1-u)_{+}$ in SVM, the exponential loss $\ell_{\text{exp}}(u)=e^{-u}$ in Adaboost, and the deviance loss $\ell_{\text{dev}}(u)=\log(1+\exp(-u))$ in logistic regression.
Given a convex surrogate loss $\ell$, the optimization problem can be formulated as
$\min_{\bm{f}\in \mathcal{F}} n^{-1}\sum_{i=1}^n\ell(M(\bm{f}(\bm{x}_i),y_i))+\lambda J(\bm{f})$,
or equivalently  $\min_{\bm{f}\in \mathcal{F}} n^{-1}\sum_{i=1}^n\ell(M(\bm{f}(\bm{x}_i),y_i))$, subject to  $J(\bm{f})\leq s_\lambda,$
where $\mathcal{F}$ is the set of candidate functions, $J(\bm{f})$ is a penalty function of $\bm{f}$, and $\lambda$ and $s_\lambda >0$ are tuning parameters.  However, solving  the above problem  is computationally heavy because of the discontinuity of the derivative of $M(\bm{f}(\bm{x}_i),y_i)$. For example, for the linear classifier  $\bm{f}(\bm{x})=\bm{A}\bm{x}$ where $\bm{A}\in \mR^{K\times p}$ and the first coordinate of $\bm{x}$ is set to 1, one can see that $\partial M(\bm{A}\bm{x}_i,y_i)/\partial \bm{A}$  is    discontinuous.
To improve the computational efficiency,  we then replace $\ell(M(\bm{f}(\bm{x}_i),y_i))$ by $V_\ell(\bm{f},\bm{z}_i)=\sum_{m=2}^{\mathcal{L}(y_i)}\sum_{\tilde{y}\in [\mathcal{E}_m(y_i)]}\ell(G_{m}(\bm{f}(\bm{x}_i),y_i,\tilde{y}))$, and estimate $\hat{\bm{f}}_\lambda$ by solving
\begin{equation}\label{V-lambda}
	\hat{\bm{f}}_\lambda=\operatorname*{argmin}\limits_{\bm{f}\in \mathcal{F}} n^{-1}\sum_{i=1}^n V_\ell(\bm{f},\bm{z}_i)+\lambda J(\bm{f}).
\end{equation}
For general loss functions, $\mathcal{F}$ is regularly chosen as the set of linear functions or the set of all measurable functions. To reduce computation, we introduce in Section 4.2 a special linear loss $\ell(u)=-u$. As shown in the Supplementary Material, a restriction $E(\|\bm{f}\|^2)\leq 1$ is required in the theoretical analysis for the linear loss. Thus, for the linear loss, $\mathcal{F}$ should be restricted on the set  $\{\bm{f}: E(\|\bm{f}\|^2)\leq 1\}$ in (\ref{V-lambda}), e.g. $\{\text{all linear functions}\}\cap \{\bm{f}: E(\|\bm{f}\|^2)\leq 1\}$ or $\{\text{all measurable functions}\}\cap \{\bm{f}: E(\|\bm{f}\|^2)\leq 1\}$. As $\lambda \rightarrow 0$, (\ref{V-lambda}) is  $\operatorname*{argmin}_{\bm{f}\in \mathcal{F}} n^{-1}\sum_{i=1}^n V_\ell(\bm{f},\bm{z}_i)$.
The theoretical properties of $\hat{\bm{f}}_\lambda$ is established in Section 5.

	\begin{remark}\label{r5}
		For the linear loss, solving (\ref{V-lambda})  might be  involved due to the restriction  $E(\|\bm{f}\|^2)\leq 1$. For easy of computation, we approximate $\hat{\bm{f}}_\lambda$ by first solving (\ref{V-lambda})  without the restriction  $E(\|\bm{f}\|^2)\leq 1$, getting an unscaled estimator,  and then scaling it to guarantee $E(\|\hat{\bm{f}}_\lambda\|^2)\leq 1$. For a large $\lambda$, the unscaled estimator can satisfy the restriction $E(\|\bm{f}\|^2)\leq 1$, that is, $\hat{\bm{f}}_\lambda$ can be calculated directly by removing the restriction $E(\|\bm{f}\|^2)\leq 1$. When $\lambda$ is small,  there may be a small gap between the actual method and the theoretical analysis. In particular, when a linear classifier is applied, %by a similar proof of (\ref{estimate_1.1}),
		one can verify that the estimator calculated in this way is tuning-parameter free and leads to the same classification result as that obtained in the next Section 4.2.
	\end{remark}
%
%
%Specifically, for the linear classifier $\bm{f}(x)=Ax$, where $A\in \mR^{K\times p}$ and the first coordinate of $x$ is set to 1, taking $J(\bm{f})$ as the Frobenius norm of $A$, we consider the optimization problem
%\begin{equation}\label{lin-min-obj}
%\min_{A\in \mR^{K\times p}}n^{-1}\sum_{i=1}^n\ell\left(M(Ax_i,y_i)\right)+\lambda \|A\|_F^2.
%\end{equation}
%\begin{equation}\label{lin-min-obj}
%\min_{A\in \mR^{K\times p}}n^{-1}\sum_{i=1}^n\ell\left(\min_{m,\tilde y\in [\mathcal{E}_m(y)]} G_{m}(Ax,y,\tilde{y})\right), \quad s.t. \quad  %\|A\|_F^2\leq \lambda.
%\end{equation}
%Considering the hinge loss function, we have
%\begin{equation}\label{svm-min}
%\min_{A\in \mR^{K\times p}}n^{-1}\sum_{i=1}^n(1-\min_{m,\tilde{y}\in [\mathcal{E}_m(y)]}G_{m}(Ax_i,y_i,\tilde{y}))_+, \quad s.t. \quad  \|A\|_F^2\leq \lambda.
%\end{equation}
%Problem (\ref{svm-min}) can be solved by dual quadratic program conventionally and the details are shown in the Supplementary Material. The tuning parameter $\lambda$ can be chosen by cross-validation.
%
%However, solving (\ref{lin-min-obj}) is still heavily  computational, since $\partial M(Ax_i,y_i)/\partial A$  is    discontinuous. We then replace $\ell(M(Ax_i,y_i))$ by $\sum_{m=2}^{\mathcal{L}(y_i)}\sum_{\tilde{y}\in [\mathcal{E}_m(y_i)]}\ell(G_{m}(Ax_i,y_i,\tilde{y}))$.
%define
%$
%L_\ell(A)=n^{-1}\sum_{i=1}^n\sum_{m=2}^{\mathcal{L}(y_i)}\sum_{\tilde{y}\in [\mathcal{E}_m(y_i)]}\ell(G_{m}(Ax_i,y_i,\tilde{y})).
%$

For the linear classifier  $\bm{f}(\bm{x})=\bm{A}\bm{x}\in\mathcal{F}$ and $J(\bm{f})$ being the square of the Frobenius norm of $\bm{A}$, we have
\begin{equation}\label{A_lambda}
	\hat {\bm{A}}_\lambda=\operatorname*{argmin}\limits_{\bm{A}} n^{-1}\sum_{i=1}^nV_\ell(\bm{A}\bm{x}_i,\bm{z}_i)+\lambda  \|\bm{A}\|_F^2.
\end{equation}
The estimated classifier is $\hat{\bm{f}}_{\lambda}(\bm{x})=\hat{ \bm{A}}_\lambda \bm{x}$.

The estimators derived from  (\ref{V-lambda}) and (\ref{A_lambda}) have two advantages. First, it is computationally efficient. %Specifically, when $\ell(\cdot)$ is taken as the linear loss introduced in the next subsection,  a closed form solution is available.
Second, as shown in Section 5.1, the population version of the estimator is Fisher consistent under mild conditions, which ensures the validity of the estimator.
Simulation and application results in Section 6 show that  our approach has obvious advantages in classification accuracy and computation, compared with other existing methods. In order to reduce computation further, in Section 4.2, we propose two specific loss functions, the linear loss and the weighted linear loss, under which a closed form of $\hat{\bm{A}}_\lambda$ can be obtained.

\subsection{Linear  loss functions}

To improve the computational efficiency of our method for massive data, we propose a linear loss function similar to that in the SVM binary classification problem \citep{Shao2015Weighted}. It leads to a closed form solution, avoiding iterations in optimization.

Define the  linear   loss function $\ell_{\text{lin}}(u)=-u$,
which is equivalent to the loss $\tilde\ell_{\mathrm{lin}}(u)=1-u$  in the sense that  both lead to the same estimator.  Note that $\ell_{\mathrm{hinge}}(u)=\max\{\tilde\ell_{\mathrm{lin}}(u),0\}$ can be viewed as the truncated version of $\tilde\ell_{\mathrm{lin}}(u)$.
Under the linear loss, we denote the unscaled minimizer of (\ref{A_lambda}) as $\hat{\bm{A}}_{\mathrm{lin},\lambda}$.  As shown in the  Supplementary Material, we have
\begin{equation}\label{estimate_1.1}
	\hat{\bm{A}}_{\mathrm{lin},\lambda}=-\bm{B}/(2\lambda),
\end{equation}
where $\bm{B}=n^{-1}\sum_{i=1}^n\sum_{m=2}^{\mathcal{L}(y_i)}\sum_{\tilde y\in [\mathcal{E}_m(y_i)]} (\bm{\xi}_m(\tilde{y})-\bm{\xi}_m(y_i))\bm{x}_i^{\top}.$
Note  that
for any two siblings  $\bm{\xi}_{j_1,j_2,\cdots,j_m}$ and $\bm{\xi}_{j_1,j_2,\cdots,j'_m}$, it holds that
$
\langle\bm{\xi}_{j_1,j_2,\cdots,j_m},\hat{\bm{A}}_{\mathrm{lin},\lambda}\bm{x}\rangle \leq \langle\bm{\xi}_{j_1,j_2,\cdots,j'_m},\hat{\bm{A}}_{\mathrm{lin},\lambda}\bm{x}\rangle \Longleftrightarrow  \langle\bm{\xi}_{j_1,j_2,\cdots,j_m},\kappa \hat{\bm{A}}_{\mathrm{lin},\lambda}\bm{x}\rangle \leq \langle\bm{\xi}_{j_1,j_2,\cdots,j'_m},\kappa \hat{\bm{A}}_{\mathrm{lin},\lambda} \bm{x}\rangle
$ for any $\kappa > 0$. Thus, there is no need to scale (\ref{estimate_1.1}) as discussed in Remark \ref{r5}.
Since the estimator $\hat{\bm{A}}_{\mathrm{lin},\lambda}$ is a linear function of $\lambda^{-1}$,
it is clear that the value of $\lambda$ does not affect the classification results.
Therefore, the estimator under the linear loss is tuning-parameter free, which can reduce  computation significantly.
We simply set $\lambda=1$ in (\ref{estimate_1.1}) and denote the estimator as $\hat{\bm{A}}_{\mathrm{lin}}$.

Although the linear loss function is simple in computation, it may not be robust to outliers. To  alleviate the impact of possible outliers, we apply the idea  of \citet{Wu:2013} and propose an adaptive weighted linear loss. We consider
%$$\hat A_{\text{ada},\lambda}=\operatorname*{argmin}\limits_{A\in \mathbb{R}^{K\times p}} n^{-1}\sum\limits_{i=1}^n w_i\left[\sum\limits_{m=2}^{\mathcal{L}(y_i)}\sum\limits_{\tilde{y}\in [\mathcal{E}_m(y_i)]}\ell_{\text{lin}}(G_{m}(Ax_i,y_i,\tilde{y}))\right]+\lambda\|A\|_F^2,$$
$$\hat{\bm{A}}_{\text{ada},\lambda}=\operatorname*{argmin}\limits_{\bm{A}} n^{-1}\sum\limits_{i=1}^n w_i V_{\ell_{\mathrm{lin}}}(\bm{A}\bm{x}_i,\bm{z}_i)+\lambda\|\bm{A}\|_F^2,$$
where $w_i$ is the adaptive weight for the $i$-th training sample.
According to \citet{Wu:2013}, we set
$w_i=1/(1+\|\hat{\bm{A}}_{\mathrm{lin}}\bm{x}_i\|^\gamma)$, where
%$\hat A_{\mathrm{lin}}$ is the estimator  of $A$ obtained by  (\ref{estimate_1.1}) with  $\lambda=1$ and
$\gamma>0$ is a tuning parameter.
One advantage of  the weighted linear loss is that the solution still has a closed form. Specifically,
$\hat{\bm{A}}_{\text{ada},\lambda}=-\bm{B}_{\text{ada}}/(2\lambda),$
where $\bm{B}_{\text{ada}}=n^{-1}\sum_{i=1}^n\sum_{m=2}^{\mathcal{L}(y_i)}\sum_{\tilde y\in [\mathcal{E}_m(y_i)]} w_i(\bm{\xi}_m(\tilde{y})-\bm{\xi}_m(y_i))\bm{x}_i^{\top}.$
The proof is similar to that of  (\ref{estimate_1.1}), and we  omit it. Clearly, the value of $\lambda$ does not affect the classification results.
We simply set $\lambda=1$ and denote the estimator as $\hat{\bm{A}}_{\text{ada}}$.
In Section 6, we see that this loss is competitive in classification accuracy and computational  efficiency.

In our simulation, we also
consider the hinge loss function $\ell_{\text{hinge}}(\cdot)$ in optimization problem (\ref{A_lambda}), which can be solved by the dual quadratic program regularly, that is,
%$$
%\min_{A\in \mR^{K\times p}}n^{-1}\sum\limits_{i=1}^n\sum\limits_{m=2}^{\mathcal{L}(y_i)}\sum\limits_{\tilde{y}\in [\mathcal{E}_m(y_i)]}(1-G_{m}(Ax_i,y_i,\tilde{y}))_++\lambda \|A\|_F^2.
%$$
$$
\min_{\bm{A}\in \mR^{K\times p}}n^{-1}\sum\limits_{i=1}^n V_{\ell_{\text{hinge}}}(\bm{A}\bm{x}_i,\bm{z}_i)+\lambda \|\bm{A}\|_F^2.
$$

\section{Statistical properties}
	
\subsection{Fisher consistency}
As stated in Section 4, the generalization error is defined based on the 0-1 hierarchical loss, that is, $
R(\bm{f})=E\left[I(Y\neq \mathcal{H}(\bm{f}(\bm{X})))\right]
$ \citep{wang2011large,babbar2016learning}.
The minimizer $\bar{\bm{f}}$ of $R(\bm{f})$ is  known as the Bayes rule.
\citet{wang2011large} proved the Bayes rule $\bar{\bm{f}}$ satisfying $\mathcal{H}(\bar{\bm{f}})=\operatorname*{argmax}_{y\in\mathcal{Y}} P(y|\bm{X}=\bm{x})$.
However, solving the optimization problem associated with the 0-1 loss is difficult.  A surrogate loss $\ell(u)$ is used instead.
For a surrogate loss $\ell(u)$, denote
%$V(\bm{f},Z)=\sum_{m=2}^{\mathcal{L}(Y)}\sum_{\widetilde Y\in [\mathcal{E}_{m}(Y)]}\ell(G_{m}(\bm{f}(X),Y,\widetilde{Y}))$ where $Z=(X,Y)$, and
%$
%R_{V}(\bm{f})=E\left[\sum_{m=2}^{\mathcal{L}(Y)}\sum_{\widetilde Y\in [\mathcal{E}_{m}(Y)]}\ell(G_{m}(\bm{f}(X),Y,\widetilde{Y}))\right].
%$
$R_{V_\ell}(\bm{f})=E[V_{\ell}(\bm{f},\bm{Z})]$ where $V_{\ell}(\bm{f},\bm{Z})=\sum_{m=2}^{\mathcal{L}(Y)}\sum_{\tilde{Y}\in [\mathcal{E}_m(Y)]}\ell(G_{m}(\bm{f}(\bm{X}),Y,\tilde{Y}))$.

Define $\bm{f}^*=\operatorname*{arginf}_{\bm{f}\in\mathcal{F}_0} R_{V_{\ell}}(\bm{f})$, where $\mathcal{F}_0$ is the set of all measurable functions. A loss $\ell(u)$ is  called  Fisher consistent, if $\bm{f}^*$ leads to the same classification rule as the Bayes rule under TD.   We establish the  Fisher consistency of  $\ell(u)$    under the following mild  conditions. Denote $P(Y^{(m)}|\bm{X}=\bm{x})$ as the conditional distribution of  the label  at the $m$-th layer, where $P(Y^{(m)}=C|\bm{X}=\bm{x})=\sum_{y\in \mathcal{Y}:y^{(m)}=C}P(y|\bm{X}=\bm{x})$.
%\citet{wang2011large} proved the Bayes rule $\bar{\bm{f}}$ respect to $R(\bm{f})$ satisfying $\mathcal{H}(\bar{\bm{f}})=\arg\max_{y\in\mathcal{Y}} P(y|x)$.
%Assuming there exists $\bar{A}\in \mathcal{F}$ such that $\mathcal{H}(\bar{A}x)=\arg\max_{y\in\mathcal{Y}} P(y|x)$, we define $\bar{A}$ as the Bayes classifier.
%To show the Fisher consistency respect to $R_{V}(\bm{f})$  under TD, we first give the following assumption.

\begin{assumption}\label{domi-path}
	Given $\bm{x}\in \mathcal{X}$, denote by  $\bar{y}=\{C_{j_1},C_{j_1,j_2},\cdots,C_{j_1,j_2,\cdots,j_{m_0}} \}$ the Bayes rule with $j_1\equiv 1$, that is, $\bar{y}=\operatorname*{argmax}_{y\in\mathcal{Y}}P(y|\bm{X}=\bm{x})$. Assume that,  for any $m=2,\cdots,m_0$,
	$$
	C_{j_1,j_2,\cdots,j_m}=\operatorname*{argmax}_{C \in\mathrm{Chi}(C_{j_1,j_2,\cdots,j_{m-1}})} P(Y^{(m)}=C|\bm{X}=\bm{x}).
	$$
\end{assumption}
Assumption \ref{domi-path} requires the conditional probability that an instance belongs to the node $C_{j_1,j_2,\cdots,j_m}$ on the path $\bar{y}$ is the largest one among siblings $\{C_{j_1,j_2,\cdots,j_m'},j'_m=1,\cdots,N_{j_1,\cdots,j_{m-1}}\}$ for $m=2,\cdots,m_0$, which is natural in hierarchical classification. Taking the example in Figure 1 as an illustration, if the Bayes classifier assigns an instance to the path $\{C_1=\{\text{Animal}\},C_{1,1}=\{\text{Elephant}\},C_{1,1,1}=\{\text{African elephant}\},C_{1,1,1,1}=\{\text{Loxodonta africana}\}\}$, Assumption \ref{domi-path} requires that the conditional probability it belongs to $C_{1,1}$ is larger than that it belongs to $C_{1,2}=\{\text{Dog}\}$, the conditional probability it belongs to $C_{1,1,1}$ is larger than that it belongs to $C_{1,1,2}=\{\text{Asian elephant}\}$, and the conditional probability it belongs to $C_{1,1,1,1}$ is larger than that it belongs to $C_{1,1,1,2}=\{\text{Loxodonta cyclotis}\}$.

\bet\label{Th3-Fish-consist}
 Under Assumption \ref{domi-path}, the loss $\ell$ is Fisher consistent by TD respect to $R_{V_{\ell}}(\bm{f})$,
 %that is, the minimizer $A^*\in\mathcal{F}$ of $R_{\ell}(A)$ satisfies $\mathcal{H}(A^*x)=\arg\max_{y\in \mathcal{Y}}P(y|x)$,
 if (i) $\ell(u)$ is differentiable with  $\ell'(u)<0$ for any $u$;  (ii)  $\ell'(u)$ is nondecreasing in $u$.
\eet
According to Theorem \ref{Th3-Fish-consist}, the exponential loss $\ell(u)=e^{-u}$, the deviance loss $\ell(u)=\log(1+\exp(-u))$, and the  linear loss $\ell(u)=-u$  satisfy the conditions in Theorem \ref{Th3-Fish-consist} and thus are Fisher consistent. Though the hinge loss  is not differentiable and the conditions above fail, \citet{Liu:2011} showed that the hinge loss is a limit of a set of large-margin unified machine loss functions, which satisfy the conditions, and then are Fisher consistent.

\begin{remark}
For a surrogate loss function $\ell(u)$,   the  minimizer of $E[\ell(M(\bm{f}(\bm{X}),Y))]$ can be  Fisher consistent without Assumption \ref{domi-path}. Details are referred to Theorem S1 in the Supplementary Material.  However, as argued in Section 4,  solving optimization problem associated with $R_{V_{\ell}}(\bm{f})$ is more computational efficient than that with  $E[\ell(M(\bm{f}(\bm{X}),Y))]$.
\end{remark}

\begin{remark}\label{r4}
	Given $y$, define $\bm{\eta}_m(y)=\bm{\eta}_{j_1,\cdots,j_m}$ if $y^{(m)}=C_{j_1,\cdots,j_m}$ for $m=2,\cdots,\mathcal{L}(y)$.
	 As shown in the Supplementary Material, for the linear loss, we have  $\bm{f}^*=\lim_{c\rightarrow +\infty}c\bm{V}_0$, where $\bm{V}_0=\sum_{y\in\mathcal{Y}}\sum_{m=2}^{\mathcal{L}(y)}P(y|\bm{x})(|\text{Sib}(\bm{\eta}_m(y))|+1)\bm{\eta}_m(y)$. In fact, all learners in the set $\{c\bm{V}_0:c>0\}$ lead to the same classification result. Taking this into account, we define $\bm{f}^*=\operatorname*{argmin}\limits_{\bm{f}\in\mathcal{F}_{0}:E(\|\bm{f}\|^2)\leq 1} R_{V_\ell}(\bm{f})=\bm{V}_0/[E(\|\bm{V}_0\|^2)]^{1/2}$ with $c=[E(\|\bm{V}_0\|^2)]^{-1/2}$.
	We prove in the Supplementary Material that the linear loss is also Fisher consistent under this definition.
\end{remark}

\subsection{Asymptotic results on the  generalization error}

In this subsection, we study the convergence rate of the excess $\ell$-risk and the excess risk \citep{Bartlett:2006}.
Recall that $\bm{f}^*$ is the underlying function that minimizes the expected loss $R_{V_{\ell}}(\bm{f})$, that is, $\bm{f}^*=\operatorname*{arginf} R_{V_{\ell}}(\bm{f})$. (For the linear loss, $\bm{f}^*$ is defined in Remark \ref{r4}.)
%where $\mathcal{F}$ is a space of  function (e.g. the space of linear learners). For simplicity, one can assume that   $\bm{f}^*\in\mathcal{F}$.
Consequently, $R_{V_{\ell}}(\bm{f}^*)$ represents the ideal performance under the surrogate loss $\ell$, whereas $R(\bm{f}^*)$ is the ideal generalization performance of $\bm{f}^*$.
Based on the Fisher consistency by Theorem \ref{Th3-Fish-consist}, we have $\mathcal{H}(\bm{f}^*)=\mathcal{H}(\bar{\bm{f}})$, and consequently $R(\bm{f}^*)=R(\bar{\bm{f}})$.
%When $\bar{\bm{f}}\notin \mathcal{F}$, $\bm{f}^*$ is an approximation of $\bar{\bm{f}}$ in the space $\mathcal{F}$.
%
%Remind that $A^*$ is the minimizer of $R_{\ell}(A)$ and $\bar{A}$ is the minimizer of $R(A)$. In Subsection 5.1, we conclude that $A^*$ is Fisher consistent under general conditions, that is, $R(A^*)=R(\bar{A})$.
The excess $\ell$-risk is then defined as $e_{V_{\ell}}(\bm{f},\bm{f}^*)=R_{V_{\ell}}(\bm{f})-R_{V_{\ell}}(\bm{f}^*)$, and the excess risk is
$e(\bm{f},\bm{f}^*)=R(\bm{f})-R(\bm{f}^*).$
Clearly, the excess $\ell$-risk measures the difference between any learning function $\bm{f}$ and $\bm{f}^*$ in terms of  the expectation $\ell$-risk, while the excess risk quantifies the difference between the hierarchical misclassification errors between $\bm{f}$ and the Bayes rule.
%{\color{blue}
%Note that $\hat{\bm{f}}_{\lambda}$ is the minimizer of (\ref{V-lambda}).
%}
For any $\dot{\bm{f}}$ and $\ddot{\bm{f}}$ in $\mathcal{F}$, let   $d(\dot{\bm{f}},\ddot{\bm{f}})=[E(\|\dot{\bm{f}}(\bm{X})-\ddot{\bm{f}}(\bm{X})\|^2)]^{1/2}$.
Remind that $\hat{\bm{f}}_\lambda$ is the minimizer of (\ref{V-lambda}), where $\mathcal{F}$ has been specified accordingly.
%Let $\mathcal{F}(\lambda)=\{\bm{f}\in\mathcal{F}: J(\bm{f})\leq s_{\lambda}\}$.
%Thus we have
%\begin{align*}
%e_{V}(\hat{\bm{f}}_{\lambda},\bm{f}^*)=R_{V}(\hat{\bm{f}}_{\lambda})-R_{V}(\bm{f}^*)=&R_{V}(\hat{\bm{f}}_{\lambda})-\inf_{\bm{f}\in\mathcal{F}(\lambda)} R_{V}(\bm{f})+\inf_{\bm{f}\in\mathcal{F}(\lambda)} R_{V}(\bm{f})-R_{V}(\bm{f}^*)
%\end{align*}
%where the second terms is the approximation error,  denoted as $\varepsilon_\lambda$.   $\varepsilon_\lambda$ will appear in the final convergence rate. In the theoretical part, we can set  $s_\lambda\ge J(\bm{f}^*)$ to get $\varepsilon_\lambda=0$.
To give asymptotic results on the generalization error, we introduce the following assumptions.

\begin{assumption}\label{excess-l-risk} Assume the loss function $\ell(u)$ is Lipschitz with a constant $0<\alpha<\infty$, that is, $
|\ell(u_1)-\ell(u_2)|\leq \alpha |u_1-u_2|
$ for any bounded $u_1$ and $u_2$.
\end{assumption}

\begin{assumption}\label{excess-risk}
	There exists constants $1\leq \gamma_1 <+\infty, 0<\gamma_2\leq +\infty$ and $\beta_i>0,i=1,2$ such that, for all small $\epsilon>0$,
	\begin{align}\label{excess-risk1}
	\inf\limits_{\{\bm{f}\in \mathcal{F}:d(\bm{f},\bm{f}^*)\geq \epsilon\}} e_{V_{\ell}}(\bm{f},\bm{f}^*)&\geq \beta_1\epsilon^{\gamma_1},\\
	\sup\limits_{\{\bm{f}\in \mathcal{F}:d(\bm{f},\bm{f}^*)\leq \epsilon\}} |e(\bm{f},\bm{f}^*)|&\leq \beta_2\epsilon^{\gamma_2}.\label{excess-risk2}
	\end{align}
\end{assumption}

Assumption \ref{excess-risk} from     \citet{Zhang:2014} enables us to control  $e(\hat{\bm{f}}_{\lambda},\bm{f}^*)$ through
$e_{V_{\ell}}(\hat{\bm{f}}_{\lambda},\bm{f}^*)$in a small neighborhood of $\bm{f}^*$.
In the Supplementary Material, we verify that under mild conditions, $\gamma_1=2$.
Note that $\gamma_2$ depends on $\bm{f}^*$ and the distribution of $\bm{Z}=(\bm{X},Y)$.
We give an illustrative example in the Supplementary Material, showing that $\gamma_2=1$ when $\mathcal{F}$ is the set of linear learners.

Before giving Assumption \ref{assum-w3}, we define a complexity measure of a function space $\mathcal{F}$. Given any $\epsilon>0$, denote $\{(\bm{f}_i^l,\bm{f}_i^u)\}$ as an $\epsilon$-bracketing function set of $\mathcal{F}$ if for any $\bm{f}\in\mathcal{F}$, there exists an $i$ such that $\bm{f}_i^l\leq \bm{f}\leq \bm{f}_i^u$ and $[E(\|\bm{f}_i^u-\bm{f}_i^l\|^2)]^{1/2}\leq \epsilon,i=1,2,\cdots$. Then the metric entropy with bracketing $\mathscr{H}_B(\epsilon,\mathcal{F})$ is the logarithm of the cardinality of the smallest $\epsilon$-bracketing set for $\mathcal{F}$. Denote $V_{\ell}^{\widetilde{T}}(\bm{f},\bm{z})=\widetilde{T}\wedge V_{\ell}(\bm{f},\bm{z})$, where $\widetilde{T}$ is a truncation constant.
Let $\bm{f}_0=\bm{f}^*$ when $\bm{f}^*\in\mathcal{F}$; otherwise, $\bm{f}_0\in\mathcal{F}$ is chosen as an approximation in $\mathcal{F}$ to $\bm{f}^*$.
Let $\mathcal{F}^{V_{\ell}}(t)=\{V_{\ell}^{\widetilde{T}}(\bm{f},\bm{z})-V_{\ell}(\bm{f}_0,\bm{z}): \bm{f}\in\mathcal{F},J(\bm{f})\leq J_0t\}$ with $J_0=\max\{J(\bm{f}_0),1\}$.

\begin{assumption}\label{assum-w3}
	For some constants $c_i>0,i=1,\cdots,3$, there exists some $\varepsilon_n>0$ such that
	$
	\sup_{t\geq 1} \phi(\varepsilon_n,t)\leq c_1n^{1/2},$
	where $\phi(\varepsilon_n,t)=\int_{c_3L}^{c_2^{1/2}L^{\beta_3/2}}\mathscr{H}_{B}^{1/2}(u,\mathcal{F}^{V_{\ell}}(t))du/L
	$
with $\beta_3=2/\gamma_1$ and  $L=L(\varepsilon_n,\lambda,t)=\min\{\varepsilon_n^2+\lambda J_0(t/2-1),1\}$.
\end{assumption}

Assumption \ref{assum-w3} measures the complexity of $\mathcal{F}^{V_{\ell}}(t)$ via the metric entropy. It was previously used in \citet{shen2007generalization,Wang:2009} and \citet{wang2011large}.
%For the metric entropy $\mathscr{H}_{B}(\epsilon,\mathcal{F}^V(t))$, when $\mathcal{F}$ is the space of linear learner, a more explicit expression is given in Lemma \ref{lemma-bracket}.
We establish  the  convergence rate  in the following theorem by the approach of  \citet{wang2011large}.
\begin{theorem}\label{excess-risk-wang}
	 Under Assumptions \ref{excess-l-risk}--\ref{assum-w3}, there exists constants $c_4$ and $c_5$ such that
	$$
	P(e(\hat{\bm{f}}_{\lambda},\bm{f}^*)\geq c_4\delta_n^{2\beta_4})\leq 3.5 \exp(-c_5n(\lambda J_0)^{2-\min(\beta_3,1)}),
	$$
	provided that $\lambda^{-1}\geq 2\delta_n^{-2}J_0$, where $\delta_n^2=\min\{\varepsilon_n^2+2e_{V_{\ell}}(\bm{f}_0,\bm{f}^*),1\}, J_0=\max\{J(\bm{f}_0),1\}, \beta_4=\gamma_2/\gamma_1$, and $\beta_3$  and $\varepsilon_n$ are defined in Assumption \ref{assum-w3}.
\end{theorem}

\begin{corollary}
	Under the assumptions in Theorem \ref{excess-risk-wang}, $|e(\hat{\bm{f}}_\lambda,\bm{f}^*)|=O_p(\delta_n^{2\beta_4})$ provided that $n(\lambda J_0)^{2-\min\{\beta_3,1\} }$ is bounded away from $0$ as $n\rightarrow \infty$.
\end{corollary}

When $\mathcal{F}$ is the set of linear functions, we have the following explicit expression  on $\mathscr{H}_{B}(\epsilon,\mathcal{F}^{V_{\ell}}(t))$.
\begin{lemma}\label{lemma-bracket}
Assume $J_0$ is bounded. Considering $\mathcal{F}$ as the set of linear functions, it holds that $\mathscr{H}_B(\epsilon,\mathcal{F}^{V_{\ell}}(t))\leq O(c_7p\log(c_6t^{1/2}/\epsilon))$, where $c_6=\max_{y\in \mathcal{Y}}\sum_{m=2}^{\mathcal{L}(y)}\left|\text{Sib}(y^{(m)})\right|T^{(m)}$ and
$c_7=\max_{y\in \mathcal{Y}}\sum_{m=2}^{\mathcal{L}(y)}\left|\text{Sib}(y^{(m)})\right|$.
\end{lemma}
%Since  the set $\{\bm{f}\in\mathcal{F},  J(\bm{f})\le J_0 t\}$ with $t\geq 1$ is already rich enough to contain $\bm{f}^*$ if $\bm{f}^*\in \mathcal{F}$, we only need to consider a bounded $t$. Assuming that both $p$ and $t$ are bounded, we have $ \mathscr{H}_B(\epsilon,\mathcal{F}^{V}(t))=O(c_7 \log(c_6/\epsilon))$.

%Note that $T^{(m)}=T^{(1)}/\delta^{m-1}$ where  $T^{(1)}$ and $\delta$ are given by ourselves in advance. As mentioned before,   we have  $\gamma_1=2$ for both the linear loss and any loss $\ell(u)$ that is twice continuously differentiable with  $\ell^{''}(u)>0$.  Moreover, $\gamma_2$ depends on the distribution of $Z=(X,Y)$ and there is no general result.
For the illustrative example in the Supplementary Material, we show that $\gamma_1=2$ and
$\gamma_2=1$ under mild conditions. Consequently, it holds that $\beta_3=1$ and $\beta_4=1/2$ there. Assuming that $p$ is bounded, we have $\mathscr{H}_B(\epsilon,\mathcal{F}^{V_{\ell}}(t))\leq O(c_7\log(c_6t^{1/2}/\epsilon))$ by Lemma \ref{lemma-bracket} when $\mathcal{F}$ is the set of linear learners.
Then by the definitions of $\phi(\varepsilon_n,t)$ and $L$,  it follows that
$\sup\limits_{t\geq 1} \phi(\varepsilon_n,t)\leq O((c_7\log(c_6/\varepsilon_n))^{1/2}/\varepsilon_n)$, and consequently $\varepsilon_n=(c_7n^{-1}\log n)^{1/2}$ by Assumption 5. Since $\bm{f}^*\in\mathcal{F}$ as shown in the Supplementary Material, the convergence rate is of the order $\sqrt{(\log n)/n}$ for this example.

For a better illustration, we compare the proposed method, named the angle-based hierarchical classification via label embedding (HierLE), with other methods by the following example of a binary tree \citep{wang2011large}. Specifically, we consider two methods, where the convergence rates have been given in literature:  (1)  multicategory SVM considering only leaf nodes (MSVM); (2) hierarchical SVM  of \cite{wang2011large} (HSVM).
%For MSVM method,  the corresponding bracketing covering number is $O(N_{nr}\log( N_{nr}/\epsilon))$ with  $N_{nr}$ is the number of non-root leafs  {\color{red} (Is it correct for $N_{nr}$?)}  \citep{shen2007generalization}. One can verify that $c_7\leq c_8\leq N_{nr}$ given that $T^{(1)}=1,\delta>1$, and that  $c_8=N_{nr}$ if and only if $k=2$ ({\color{red}??? Correct?}), that is, there is no hierarchical structure. When $k>2$ and structure of the tree is complicated,  $\mathscr{H}_B(\epsilon,\mathcal{F}^{V}(t))$ can be much smaller than $N_{nr}\log( N_{nr}/\epsilon)$.
%For the HSVM considered  in \citet{wang2011large}, the corresponding bracketing covering number is $O(2c(\mathcal{H})\log(2c(\mathcal{H})/\epsilon))$ where $c(\mathcal{H})=\sum_{j=0}^q|\text{Chi}(C_{(j)})|(|\text{Chi}(C_{(j)})|-1)/2$, {\color{red}where $\{C_{(q)}, q=0,\cdots, q\}$ is the set of all non-leaf nodes of the tree.(Is it correct?)}
%One can verify that $c_7\leq c_8\leq 2c(\mathcal{H})$ given that $T^{(1)}=1$ and $\delta>1$. When the tree structure is complicated, $c_8$ can be  much smaller than $2c(\mathcal{H})$.

{\bf Example.} For a binary tree with depth $k$, it has  $n_{\mathrm{leaf}}=2^{k-1}$ leaf nodes and $2^k-2$ non-root nodes.
Recall $\mathcal{C}_{H}=\{C_{(1)},\cdots,C_{(q)}\}$ is the set of non-root nodes, and $Y$ is a path from the root to a leaf node.
Denote the set of leaf nodes  as $\mathcal{C}_{\text{leaf}}=\{\widetilde{C}_{(i)}, i=1,\cdots, 2^{k-1}\}\subset \mathcal{C}_{H}$. The conditional probability of  $Y=y$ can be expressed as $P(Y=y|\bm{X}=\bm{x})=P(y\cap \mathcal{C}_{\text{leaf}}|\bm{X}=\bm{x})$. Let $\bm{X}\in\mR^2$ sampled from the uniform distribution over $[0,1]^2$. Given $\bm{x}=(x_1,x_2)^\top$, when  $x_1\in [(i-1)/2^{k-1},i/2^{k-1}), i=1,\cdots, 2^{k-1}$, define the probability
$P(\widetilde{C}_{(i)}\in Y|\bm{X}=\bm{x})=1-2^{-(k-1)}$,  and for any $\widetilde{C}_{(j)}\in \mathcal{C}_{\text{leaf}}$ with $j\ne i$, let
$ P(\widetilde{C}_{(j)}\in Y|\bm{X}=\bm{x})\equiv 2^{-(k-1)}/(2^{k-1}-1)$.
Moreover,  for any non-leaf node $C_{(i)}\in \mathcal{C}_H\backslash \mathcal{C}_{\text{leaf}}$, let $P(C_{(i)}\in Y|\bm{X}=\bm{x})=\sum_{t\in\text{Off}(C_{(i)}) \cap \mathcal{C}_{\text{leaf}}} P(t\in Y|\bm{X}=\bm{x})$.  %The Bayes rule $\bar{\bm{f}}$ predicts $X$  to the path from the root to the leaf node $\widetilde{C}_{(i)}$ when $X_1\in [(i-1)/2^{k-1},i/2^{k-1}), i=1,\cdots, 2^{k-1}$.})
We consider the linear classifiers for this example. One can verify the optimal minimizer is $\bm{f}^*(\bm{x})=(f_1(\bm{x}),f_2(\bm{x}),\cdots,f_K(\bm{x}))^\top\in \mathcal{F}$ with $f_1(\bm{x})=x_1-2^{k-2}/2^{k-1}$, $f_j(\bm{x})=f_{\lfloor j/2 \rfloor}+[2I(j\mod 2=0)-1](x_1-f_{\lfloor j/2 \rfloor})/2$ for $j=2,\cdots,K$.

Let us consider our HierLE estimator first.
By \citet{wang2011large}, we have  $\beta_3=1$ and  $\beta_4=1/2$.
%To obtain the   convergence rate of HierLE  for the above example,  we then derive the   expression of $\varepsilon_n$ in Assumption 5.
%Recall  that $\phi(\tilde{\varepsilon}_n,t)$ %=\int_{c_5L}^{c_4^{1/2}L^{\beta_4/2}}\mathscr{H}_{B}^{1/2}(u,\mathcal{F}^V(t))du/L$
% and that  $L=L(\tilde{\varepsilon}_n,\lambda,t)=\min\{\tilde{\varepsilon}_n^2+\lambda J_0(t/2-1),1\}$ in Assumption 5.
Recall   that $\mathscr{H}_B(\epsilon,\mathcal{F}^{V_{\ell}}(t))\leq O(c_7\log(c_6t^{1/2}/\epsilon))$ in Lemma \ref{lemma-bracket}. Then by the definitions of $\phi(\varepsilon_n,t)$ and $L$,    it follows that
$\sup\limits_{t\geq 1} \phi(\varepsilon_n,t)\leq O((c_7\log(c_6/\varepsilon_n))^{1/2}/\varepsilon_n)$, and consequently that
$\varepsilon_n=(c_7n^{-1}\log n)^{1/2}$ by Assumption 5. Moreover, for this example, one can compute that $c_7=k-1$.  By  Theorem  \ref{excess-risk-wang} and Corollary 1, we have $|e(\hat{\bm{f}}_\lambda,\bm{f}^*)|=O_p(\varepsilon_n)=O_p((kn^{-1}\log n)^{1/2})$.

For comparisons, we consider the rates of  MSVM and HSVM for this example.   The convergence rate of  MSVM is $O_p((n_{\text{leaf}}(n_{\text{leaf}}-1)n^{-1}\log n/2)^{1/2})$ \citep{wang2011large}.  For any given tree, define  $c_8=\sum_{j=0}^q|\text{Chi}(C_{(j)})|(|\text{Chi}(C_{(j)})|-1)/2$, where $\{C_{(j)}, q=0,\cdots, q\}$ is the set of all nodes one the tree.  For the binary tree above, we have $c_8=2^{k-1}-1=n_{\mathrm{leaf}}-1$.  By \citet{wang2011large}, the HSVM has the convergence rate   $O_p((n_{\mathrm{leaf}} n^{-1}\log n)^{1/2})$.
Clearly,  $c_7$ is  much smaller than $n_{\mathrm{leaf}}$ and $n_{\mathrm{leaf}}(n_{\mathrm{leaf}}-1)/2$ for  this example. Therefore, the proposed estimator has advantages  compared with these two existing methods.

For  better illustration,   we check   the candidate set  $\mathcal{F}$ when the linear classifier is applied.   For our method, the candidate set of linear classifiers is  $\mathcal{F}=\{\bm{f}:\bm{f}=\bm{A}\bm{x}\in\mathbb{R}^{K},\bm{A}\in\mathbb{R}^{K\times p},\bm{x}\in\mathbb{R}^p\}$, where the intercept is included in $\bm{x}$.  For  HSVM, the candidate set is $\mathcal{F}=\{\bm{f}:\bm{f}=\bm{W}\bm{x}\in\mathbb{R}^q,\bm{W}\in\mathbb{R}^{q\times p},\bm{x}\in\mathbb{R}^{p}\}$ \citep{wang2011large}. It is seen that the candidate set $\mathcal{F}$ for our method is related to   the dimension $K$ of embedding space, while $\mathcal{F}$ for HSVM is associated with $q$, the total number of nodes except the root. As shown in Section 3, $K$ is  smaller than $q$.
On the other hand,  Assumption \ref{domi-path} is required for our method to obtain Fisher consistency. For HSVM, Fisher consistency can be established in more general situations without Assumption \ref{domi-path}.

\section{Simulation and real data analysis}
In this section, we evaluate the  proposed method under three loss functions denoted as HierLE$_{\text{lin}}$ (linear loss), HierLE$_{\text{wl}}$ (weighted linear loss) and HierLE$_{\text{hinge}}$ (hinge loss), and compare them  with some competitors. Specifically, we consider (1) MSVM \citep{Chang:2011}; %(multicategory SVM considering  binary classification for all pairs of leaf nodes ),
(2) sequential hierarchical SVM training SVMs separately for each parent node and using the top-down strategy to assign labels \citep{davies2007hierarchical} (SHSVM);
(3) sequential hierarchical binary SVM training binary SVMs separately for each node and using the top-down strategy to assign labels \citep{cesa2006hierarchical} (SBSVM);
(4) traditional label embedding method for hierarchical classification \citep{Cai:2004} (HofSVM); (5) HSVM \citep{wang2011large};
(6) cost-sensitive learning method for  hierarchical classification that can be used for massive data \citep{Charuvaka:2015} (HierCost).

\subsection{Evaluation measures}
Given the test set $\{(\bm{x}_i,y_i)\}_{i=1}^{n_{\textrm{te}}}$  of size $n_{\textrm{te}}$, denote $\hat y_i=\{\hat y_i^{(1)},\cdots, \hat y_i^{(\mathcal{L}(\hat{y}_i))}\}$ as the estimated path of $\bm{x}_i$, $i=1,\cdots,n_{\textrm{te}}$.
We introduce first four losses and then the hierarchical f-measure suggested by \cite{Silla:2011}. Note that smaller values are preferred for  the four losses, and larger values are preferred for hF.

The first evaluation metric is the 0-1 loss \citep{Cai:2004}, which gives loss of 0 if $\bm{x}_i$ is labeled correctly in the whole path, and 1 otherwise, that is,
$	\ell_{0-1}=\sum_{i=1}^{n_{\text{te}}}I(\hat{y_i}\neq y_i)/n_{\text{te}}.
$
The symmetric loss $\ell_{\Delta}$ is calculated as follows \citep{kosmopoulos2015evaluation},
$
\ell_{\Delta}=\sum_{i=1}^{n_{\text{te}}}|(\hat{y}_i\backslash y_i)\cup (y_i\backslash \hat{y}_i)|/n_{\text{te}}.
$
The symmetric loss treats each node on the tree equally. \citet{Cesa-Bianchi:2004}  defined a hierarchical loss function which  views the  mistakes made at higher layers being more important than those  at lower layers.
%As shown in Subsection 2.1, we sort nodes in $\mathcal{C}_H$ by layers from top to bottom, and the nodes at the same layer from left to right. Then denote these sorted nodes as $\{1,2,\cdots, q\}$, where $q=|\mathcal{C}_H|$.
Note that $y_i$ and $\hat y_i$ may have different lengths. We transform $y_i$ into a binary vector $\bm{\mathcal{Q}}(y_i)\in \mR^q$, where the $j$-th coordinate $\mathcal{Q}(y_i)_j$ indicates whether the node $C_{(j)}$ defined in Section 2.1 is on the path $y_i$. Define $\bm{\mathcal{Q}}(\hat y_i)$ in the same way. The hierarchical loss is calculated as
$
\ell_\text{H}=\sum_{i=1}^{n_{\textrm{te}}}\sum_{j=1}^{q}v_{C_{(j)}}I(\{\mathcal{Q}(\hat{y}_i)_j\neq \mathcal{Q}(y_i)_j\} \wedge \{\mathcal{Q}(\hat{y}_i)_s=\mathcal{Q}(y_i)_s,\ \forall s< j\})/n_{\text{te}}.
$
The coefficients $0 \leq v_{C_{(j)}}\leq 1$ are used for down-scaling the loss. There are two popular choices  for $v_{C_{(j)}}$. Specifically, denote $\ell_\mathrm{H}$ as $\ell_{\text{H(sib)}}$ when  $v_{C_{(j)}}$ takes the form
$
v_{C_{(0)}}=1,v_{C_{(j)}}=v_{\text{Par}(C_{(j)})}/|\text{Sib}(C_{(j)})|,j=1,\cdots,q,
$
where  $|\text{Sib}(C_{(j)})|$ represents the number of siblings of the node $C_{(j)}$. Denote  $\ell_\mathrm{H}$ as $\ell_{\text{H(sub)}}$ when $v_{C_{(j)}}$ is of the form
$
v_{C_{(j)}}= q^{-1}|\text{subtree}(C_{(j)})|, j=1,\cdots,q,
$
where $|\text{subtree}(C_{(j)})|$ is the size of the subtree rooted by the node $C_{(j)}$.

Besides the four losses above, 	\citet{Silla:2011} suggested using the hierarchical f-measure \citep{Kiritchenko:2005}, as it can be effectively
applied to any hierarchical classification
scenario, i.e. tree, DAG, single-labeled and multiple-labeled. It is defined as
$
\text{hF}=2\cdot\text{hP}\cdot\text{hR}/(\text{hP}+\text{hR}),
$
where hP and hR are hierarchical precision and hierarchical recall, respectively defined as,
\begin{align*}
\text{hP}=&\frac{\sum_{i=1}^{n_{\text{te}}}|
\{\cup_{C_{(j)}\in \hat{y}_i}\text{Anc}(C_{(j)})\cup \hat{y}_i\}\cap \{\cup_{C_{(j)}\in y_i}\text{Anc}(C_{(j)})\cup y_i\}|}{\sum_{i=1}^{n_{\text{te}}}|\cup_{C_{(j)}\in \hat{y}_i}\text{Anc}(C_{(j)})\cup \hat{y}_i|}\\ \text{hR}=&\frac{\sum_{i=1}^{n_{\text{te}}}|\{\cup_{C_{(j)}\in \hat{y}_i}\text{Anc}(C_{(j)})\cup \hat{y}_i\}\cap \{\cup_{C_{(j)}\in y_i}\text{Anc}(C_{(j)})\cup y_i\}|}{\sum_{i=1}^{n_{\text{te}}}|\cup_{C_{(j)}\in y_i}\text{Anc}(C_{(j)})\cup y_i|}.
\end{align*}

\subsection{Simulation}
In our simulations, samples are split into the training, the validation and the test sets, with sizes denoted as $n, n_{\text{vl}}$ and  $n_{\text{te}}$, respectively.
We set $n:n_{\text{vl}}:n_{\text{te}}=1:1:2$. Let $T^{(1)}=1$ and $\delta=\sqrt{5}$ in Algorithm 2 as discussed in Remark 2 in  Section 3. We first learn classifiers on the training set and choose the best tuning  parameter based on the validation set over 41 grid points $\{10^{i/10},i=-20,-19,\cdots,20\}$.
Note that HierLE$_{\text{lin}}$ is tuning free.
%For   any method that needs to choose tuning parameters, %{\color{red}including FlatSVM, HofSVM, ParSVM, Hiercost, TDWL and TDSVM,}
%we choose the best one using  the validation set by a grid search over points $\{10^{-2+i/10},i=0,\cdots,30\}$ in the interval  $[10^{-2},10]$.  %\{10^{-2+i/10},i=0,\cdots,30\}$.
Then we apply the estimated learner on the test set to compute the evaluation metrics.

\textbf{Example 1}.
We first consider a tree of $k$ layers. There are 4 nodes at the second layer and each node has two children at the lower layers. The tree structure is shown in Figure \ref{fig:stru_s} (left), where the digits stand for labels.
Note that all leaf nodes locate at the $k$-th layer.
Simulate data $\{(\bm{x}_i,y_i)\}_{i=1}^{4n}$ as follows, where $\bm{x}_i=(x_{i1},\cdots,x_{ip})^\top$, $y_i=\{y_{i}^{(1)},\cdots,y_{i}^{(k)}\}^\top$. For $i=1,2,\cdots,4n$, $y_i$ follows a discrete uniform distribution in the set of paths on the tree. Let $\bm{\mu}_i$ be a zero vector of length $p$ except for the $y_i^{(m)}$-th element being $1/(m-1)$ for $m=2,\cdots,k$. Taking $k=3,y_i=\{0,1,5\}$ as an example, we have $\bm{\mu}_i=(1,0,0,0,1/2,0,\cdots,0)^\top$.
Let $\bm{x}_i|y_i \sim N(\bm{\mu}_i,0.1\bm{I}_{p\times p})$. Then we randomly select 20\% of the data and assign
their labels randomly to generate non-separable cases.
Set $n=50$, $k=3,4$ and $p=15,30$.
\begin{figure}[h]
	\centering
	\begin{minipage}[b]{0.49\textwidth}
		\centering
		\includegraphics[width=7.8cm]{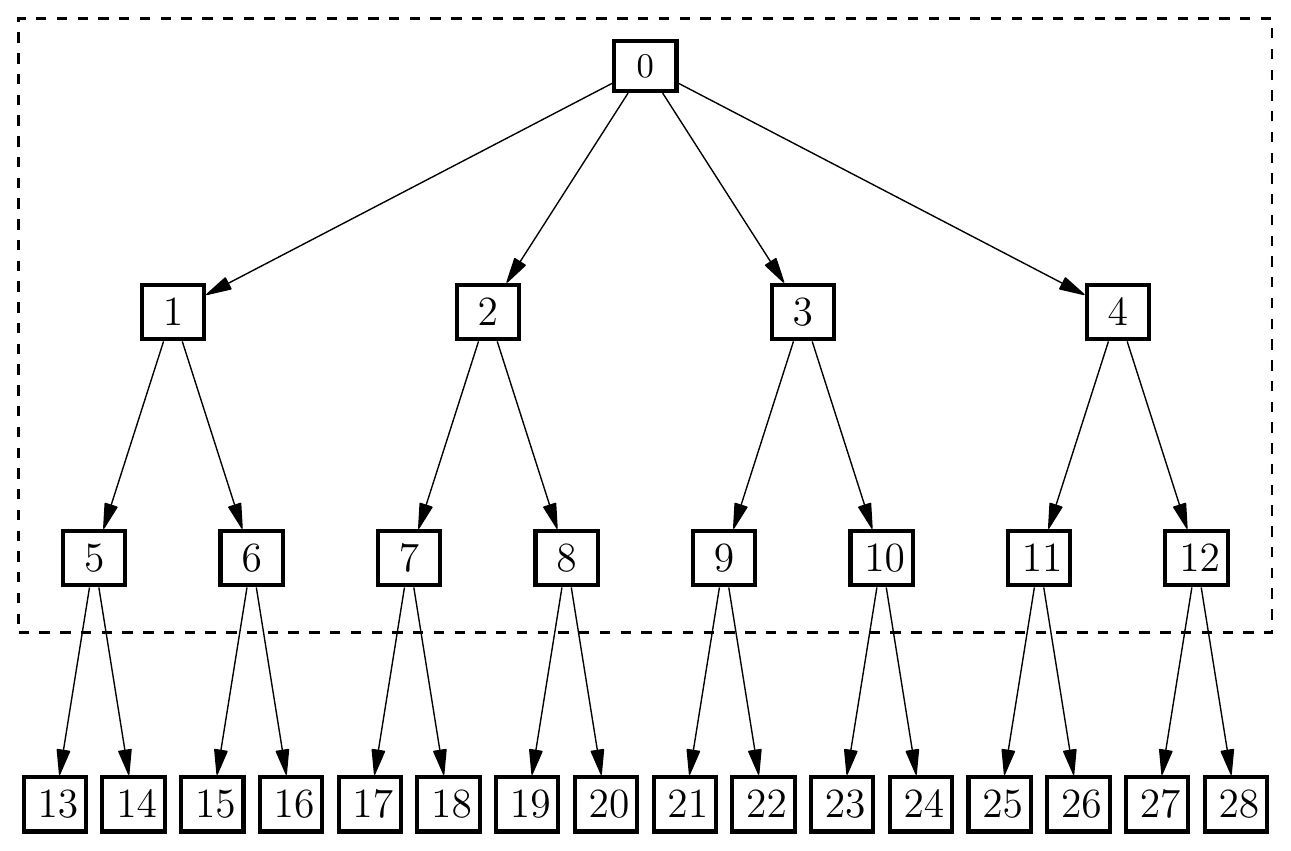}
	\end{minipage}
	\begin{minipage}[b]{0.49\textwidth}
		\centering
		\includegraphics[width=7.8cm]{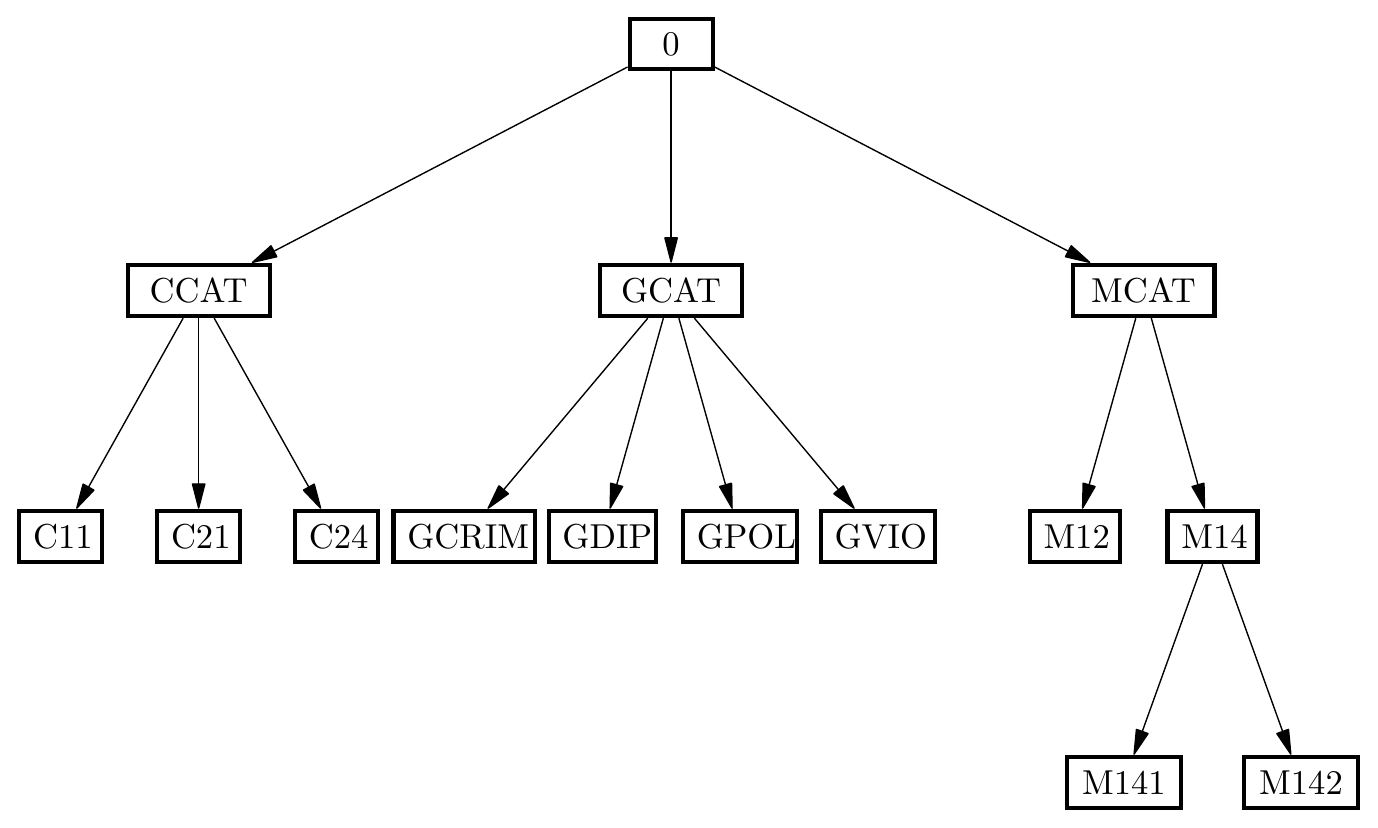}
	\end{minipage}
	\caption{The hierarchical structure for Example 1 (left) and the subtree of Reuters (right).}
	\label{fig:stru_s}
\end{figure}

The average results over 100 replications on $\ell_{0-1}, \ell_{\Delta}, \ell_{\text{H}(sib)}, \ell_{\text{H(sub)}},\text{hF}$ and  time cost (i.e. time on training, validating  and testing) are shown in Table \ref{tab:simu1}.
From Table 1, we see that in terms of  the four loss measures   and the hF measure, SBSVM performs worst. Our proposed three classifiers HierLE$_{\text{lin}}$, HierLE$_{\text{wl}}$ and HierLE$_{\text{hinge}}$  perform  better than other methods with HierLE$_{\text{wl}}$ being  the  best in all evaluation metrics. Compared to MSVM, all hierarchical classifiers get better as $k$ increases.
Regarding the computational time, HofSVM takes the longest to run, followed by HSVM, HierCost, HierLE$_{\text{hinge}}$, SBSVM, MSVM,  SHSVM, HierLE$_{\text{wl}}$ and HierLE$_{\text{lin}}$. Our proposed method runs fast under the (weighted) linear loss since it has a closed form solution.

\begin{sidewaystable}[htbp]
	\centering
	\caption{Average results as well as standard deviations for Example 1 and Example 2 over 100 replications. The best value in each column is boldfaced and the percentages in brackets are the amounts of improvement over MSVM.}
	\begin{tabular}{p{2cm}<{\raggedright}p{3.3cm}<{\raggedright}p{3.2cm}<{\raggedright}p{3.2cm}<{\raggedright}p{3.2cm}<{\raggedright}p{3.3cm}<{\raggedright}p{1cm}<{\raggedright}}
		\toprule
		& $\ell_{0-1}$& $\ell_{\Delta}$& $\ell_{\text{H(sib)}}$& $\ell_{\text{H(sub)}}$& hF&Time\\
		\midrule
		& \multicolumn{6}{c}{{\bf Example 1} $k=3$, $p=15$} \\
		MSVM     & 0.541$_{0.006}$ (0.0\%)          & 1.716$_{0.023}$(0.0\%)          & 0.107$_{0.001}$(0.0\%)          & 0.098$_{0.001}$(0.0\%)          & 0.571$_{0.006}$(0.0\%)          & 0.510    \\
		SHSVM    & 0.521$_{0.007}$(3.7\%)          & 1.595$_{0.023}$(7.0\%)          & 0.100$_{0.001}$(7.0\%)          & 0.090$_{0.001}$(9.1\%)          & 0.601$_{0.006}$(5.3\%)          & 0.290    \\
		SBSVM    & 0.589$_{0.006}$(-8.9\%)          & 1.910$_{0.023}$(-11.3\%)          & 0.119$_{0.001}$(-11.3\%)          & 0.110$_{0.001}$(-11.8\%)          & 0.522$_{0.006}$(-8.5\%)          & 0.777   \\
		HofSVM  & 0.541$_{0.006}$( 0.0\%)          & 1.744$_{0.022}$(-1.7\%)          & 0.109$_{0.001}$(-1.7\%)          & 0.101$_{0.001}$(-2.6\%)          & 0.564$_{0.006}$(-1.2\%)          & 31.421  \\
		HSVM     & 0.517$_{0.006}$(4.5\%)          & 1.627$_{0.022}$(5.2\%)          & 0.102$_{0.001}$(5.2\%)          & 0.093$_{0.001}$(5.7\%)          & 0.593$_{0.005}$(3.9\%)          & 6.580    \\
		HierCost & 0.517$_{0.006}$(4.4\%)          & 1.635$_{0.023}$(4.7\%)          & 0.102$_{0.001}$(4.7\%)          & 0.092$_{0.001}$(6.3\%)          & 0.591$_{0.006}$(3.5\%)          & 6.589   \\
		HierLE$_{\text{lin}}$     & 0.501$_{0.006}$(7.4\%)          & 1.518$_{0.019}$(11.5\%)          & 0.095$_{0.001}$(11.5\%)          & 0.085$_{0.001}$(13.9\%)          & 0.620$_{0.005}$(8.6\%)          & 0.014   \\
		HierLE$_{\text{wl}}$    & \textbf{0.498$_{0.006}$(8.0\%)} & \textbf{1.495$_{0.018}$(12.9\%)} & \textbf{0.093$_{0.001}$(12.9\%)} & \textbf{0.083$_{0.001}$(15.7\%)} & \textbf{0.626$_{0.004}$(9.7\%)} & 0.553   \\
		HierLE$_{\text{hinge}}$     & 0.507$_{0.006}$(6.2\%)          & 1.515$_{0.021}$(11.7\%)          & 0.095$_{0.001}$(11.7\%)          & 0.083$_{0.001}$(15.5\%)          & 0.621$_{0.005}$(8.8\%)          & 6.017   \\
		& \multicolumn{6}{c}{{\bf Example 1} $k=4$, $p=30$}                                                                                                                                                 \\
		MSVM     & 0.791$_{0.005}$(0.0\%)           & 3.684$_{0.034}$(0.0\%)           & 0.142$_{0.001}$(0.0\%)           & 0.134$_{0.001}$(0.0\%)           & 0.386$_{0.006}$(0.0\%)           & 0.807   \\
		SHSVM    & 0.770$_{0.005}$(2.7\%)           & 3.398$_{0.031}$(7.8\%)           & 0.128$_{0.001}$(9.8\%)           & 0.119$_{0.001}$(11.5\%)           & 0.434$_{0.005}$(12.4\%)           & 1.023   \\
		SBSVM    & 0.816$_{0.005}$(-3.2\%)           & 3.849$_{0.034}$(-4.5\%)           & 0.149$_{0.001}$(-4.6\%)           & 0.141$_{0.002}$(-4.7\%)           & 0.358$_{0.006}$(-7.1\%)           & 2.287   \\
		HofSVM  & 0.779$_{0.005}$(1.6\%)           & 3.677$_{0.033}$(0.2\%)           & 0.143$_{0.001}$(-0.5\%)           & 0.136$_{0.002}$(-0.9\%)           & 0.387$_{0.006}$(0.3\%)           & 337.014 \\
		HSVM     & 0.753$_{0.005}$(4.9\%)           & 3.401$_{0.034}$(7.7\%)           & 0.130$_{0.001}$(8.8\%)           & 0.121$_{0.001}$(9.8\%)           & 0.433$_{0.006}$(12.2\%)           & 29.643  \\
		HierCost & 0.784$_{0.005}$(0.9\%)           & 3.563$_{0.032}$(3.3\%)           & 0.137$_{0.001}$(3.7\%)           & 0.129$_{0.001}$(4.2\%)           & 0.406$_{0.005}$(5.2\%)           & 18.799  \\
		HierLE$_{\text{lin}}$     & \textbf{0.735$_{0.005}$(7.2\%)}  & 3.134$_{0.028}$(14.9\%)           & 0.117$_{0.001}$(17.5\%)           & 0.108$_{0.001}$(19.9\%)           & 0.478$_{0.005}$(23.8\%)           & 0.022   \\
		HierLE$_{\text{wl}}$     & \textbf{0.735$_{0.005}$(7.2\%)}  & \textbf{3.130$_{0.028}$(15.0\%)}  & \textbf{0.117$_{0.001}$(17.7\%)}  & \textbf{0.107$_{0.001}$(20.1\%)}  & \textbf{0.478$_{0.005}$(23.9\%)}  & 0.842   \\
		HierLE$_{\text{hinge}}$     & 0.751$_{0.005}$(5.2\%)           & 3.148$_{0.028}$(14.5\%)           & 0.118$_{0.001}$(17.2\%)           & 0.108$_{0.001}$(19.6\%)           & 0.475$_{0.005}$(23.1\%)           & 9.890\\
		 & \multicolumn{6}{c}{{\bf Example 2}}                                                                                                                                                     \\
		MSVM     & 0.814$_{0.001}$(0.0\%)          & 4.288$_{0.005}$(0.0\%)          & 0.036$_{0.000}$(0.0\%)          & 0.033$_{0.000}$(0.0\%)          & 0.464$_{0.001}$(0.0\%)          & 258.470   \\
		SHSVM    & 0.799$_{0.001}$(1.9\%)          & 4.001$_{0.005}$(6.7\%)          & 0.032$_{0.000}$(10.3\%)          & 0.030$_{0.000}$(11.4\%)          & 0.500$_{0.001}$(7.8\%)          & 105.416  \\
		SBSVM    & 0.835$_{0.001}$(-2.5\%)          & 4.690$_{0.006}$(-9.4\%)          & 0.041$_{0.000}$(-14.2\%)          & 0.039$_{0.000}$(-15.8\%)          & 0.414$_{0.001}$(-10.8\%)          & 1452.514 \\
		HierCost & 0.802$_{0.001}$(1.5\%)          & 4.098$_{0.010}$(4.4\%)          & 0.033$_{0.000}$(6.3\%)          & 0.031$_{0.000}$(7.0\%)          & 0.488$_{0.001}$(5.1\%)          & 575.524  \\
		HierLE$_{\text{lin}}$     & 0.790$_{0.001}$(3.1\%)          & 3.781$_{0.005}$(11.8\%)          & 0.029$_{0.000}$(17.9\%)          & 0.027$_{0.000}$(19.9\%)          & 0.527$_{0.001}$(13.7\%)          & 1.871    \\
		HierLE$_{\text{wl}}$     & \textbf{0.788$_{0.001}$(3.2\%)} & \textbf{3.742$_{0.005}$(12.7\%)} & \textbf{0.029$_{0.000}$(19.2\%)} & \textbf{0.026$_{0.000}$(21.4\%)} & \textbf{0.532$_{0.001}$(14.7\%)} & 58.129\\
		\bottomrule
	\end{tabular}
	\label{tab:simu1}%
\end{sidewaystable}%

\textbf{Example 2}.
In this example, we simulate a more complex tree and samples of a larger size. The hierarchy is of 5 layers.
There are 12 nodes at the second layer and each node has two children at the lower layers. Thus, there are 24, 48, 96 nodes at the third, fourth and fifth layers, respectively, and 180 nodes except for the root in total.
Note that all leaf nodes locate at the last layer.
We simulate $\{(\bm{x}_i,y_i)\}_{i=1}^{4n}$ as follows, where $\bm{x}_i=(x_{i1},\cdots,x_{ip})^\top,y_i=\{y_{i}^{(1)},y_{i}^{(2)},y_{i}^{(3)},y_{i}^{(4)},y_{i}^{(5)}\}^\top$. For $i=1,2,\cdots,4n$,  $y_i$ follows a discrete uniform distribution in the set of paths on the tree.
For each $\bm{x}_i$, $\bm{x}_i|y_i\sim N(\bm{\xi}_5(y_i),0.1\bm{I}_{p\times p})$, where $\bm{\xi}_5(y_i)$ is the constructed point corresponding to the leaf node $y_i^{(5)}$ and $p=n_{\text{leaf}}-1$. Set $n=2000$.

Since the size of this problem is large,
HofSVM, HSVM and HierLE$_{\text{hinge}}$,  involving quadratic programming with large amounts of constraints,    are not considered because of computational inefficiency. The average results  for  MSVM, SHSVM, SBSVM, HierCost, HierLE$_{\text{lin}}$ and HierLE$_{\text{wl}}$ over 100 replications  are shown in Table \ref{tab:simu1}.
It is seen  that the  proposed classifiers HierLE$_{\text{lin}}$ and HierLE$_{\text{wl}}$ have significant advantages over other methods in terms of both classification accuracy and computing costs for this large dataset.

\subsection{Real data analysis: document categorization}

Hierarchical classification has a wide range of applications in text categorization, particularly on the Web.
As the complexity of the hierarchy and the number of documents increase, it is desirable to consider methods that incorporate the hierarchical information and that can be computed efficiently.

The first dataset is a part of Reuters \citep{lewis2004rcv1}, which is an archive of manually categorized newswire
stories$^1$\footnote{$^1$The dataset  is available at http://kt.ijs.si/DragiKocev/PhD/resources/doku.php?id=hmc\_classification}. It is a multi-labeled (one sample may belong to several paths)  hierarchical classification problem. There are 3000 samples in the original data set with 47236 features.
From the whole tree, we select a subtree as shown in Figure \ref{fig:stru_s} (right). It has four layers with 15 nodes in total and 10 leaf nodes.  Then we select observations that only belong to one of the paths in our selected subtree as our samples. The sample size is 455 and the number of features is 7206 for this new small dataset.

The second dataset Chinese Hierarchical Text Classification (CHTC) is collected by the authors. We download  $5241$ advertisements from an online shopping website in China. The categories are organized in a hierarchical tree of $4$ layers. There are 5 nodes at the second layer, indicating whether the advertisement belongs to food, amusements, life services, online shopping, or travel. There are 16 and 35 nodes at the third and fourth layers, respectively. Totally, there are 57 nodes with 40 leaf nodes. Detailed information is shown in the Supplementary Material. We perform documents parsing and tokenization to get $13103$ terms. The covariates represent the frequency of these terms.
For this dataset, we consider two cases, the first 3 layers of the whole tree and the whole tree itself.

For each dataset, we assign the sample into  the training, validation and test sets with ratio 1:1:2.  We perform feature screening via distance correlation \citep{Li:2012} to select 110 important features for Reuters and 1000 important features for the two cases of CHTC.

The average results over 100 replications are shown in Table \ref{tab:CHTC}.
For the small subtree of Reuters, we compare all nine methods. Our methods HierLE$_\text{lin}$, HierLE$_\text{wl}$ and HierLE$_\text{hinge}$ are better than other methods in all evaluation measures and HierLE$_\text{wl}$ performs best.
Regarding the computational time, the proposed methods  HierLE$_\text{lin}$ and HierLE$_\text{wl}$ are quite efficient, with HierLE$_\text{lin}$ being the best one.
For the large tree with $k=3,4$ of CHTC, we compare only MSVM, SHSVM, SBSVM, HierCost, HierLE$_\text{lin}$ and HierLE$_\text{wl}$ because the other three methods are very time consuming. We can see that HierCost, HierLE$_\text{lin}$ and HierLE$_\text{wl}$ perform better than MSVM in all metrics, while SHSVM and SBSVM are worse. Moreover, our proposed methods HierLE$_\text{lin}$ and HierLE$_\text{wl}$ show great advantages in terms of classification accuracy and computational efficiency.

\begin{sidewaystable}
	\centering
	\caption{Average results as well as standard deviations for Reuters and CHTC over 100 replications. The best value in each column is boldfaced and the percentages in brackets are the amounts of improvement over MSVM.}
	\small
	\begin{tabular}{p{2cm}<{\raggedright}p{3.4cm}<{\raggedright}p{3.4cm}<{\raggedright}p{3.4cm}<{\raggedright}p{3.4cm}<{\raggedright}p{3.4cm}<{\raggedright}p{1cm}<{\raggedright}}
	\toprule
	& $\ell_{0-1}$& $\ell_{\Delta}$& $\ell_{\text{H(sib)}}$& $\ell_{\text{H(sub)}}$& hF&Time\\
	\midrule
	& \multicolumn{6}{c}{\textbf{Reuters}} \\
	MSVM & 0.435$_{0.004}$(  0.0\%) & 1.298$_{0.014}$(  0.0\%) & 0.079$_{0.001}$(  0.0\%) & 0.066$_{0.001}$(  0.0\%) & 0.724$_{0.003}$(  0.0\%) & 1.893 \\
	SHSVM & 0.430$_{0.003}$(  1.1\%) & 1.225$_{0.012}$(  5.6\%) & 0.073$_{0.001}$(  7.7\%) & 0.060$_{0.001}$(  9.4\%) & 0.738$_{0.003}$(  2.0\%) & 1.236 \\
	SBSVM & 0.479$_{0.004}$(-10.0\%) & 1.361$_{0.012}$( -4.8\%) & 0.083$_{0.001}$( -4.1\%) & 0.068$_{0.001}$( -3.1\%) & 0.708$_{0.003}$( -2.2\%) & 3.196 \\
	HofSVM & 0.435$_{0.004}$(  0.2\%) & 1.277$_{0.012}$(  1.6\%) & 0.079$_{0.001}$(  0.2\%) & 0.065$_{0.001}$(  0.9\%) & 0.724$_{0.003}$(  0.1\%) & 644.632 \\
	HSVM & 0.431$_{0.004}$(  1.1\%) & 1.196$_{0.011}$(  7.8\%) & 0.072$_{0.001}$(  9.0\%) & 0.058$_{0.001}$( 12.4\%) & 0.741$_{0.002}$(  2.4\%) & 55.016 \\
	HierCost & 0.475$_{0.004}$( -9.2\%) & 1.423$_{0.014}$( -9.6\%) & 0.089$_{0.001}$(-11.9\%) & 0.075$_{0.001}$(-14.1\%) & 0.691$_{0.003}$( -4.6\%) & 12.272 \\
	HierLE$_{\text{lin}}$ & 0.428$_{0.004}$(  1.6\%) & 1.161$_{0.013}$( 10.6\%) & 0.068$_{0.001}$( 13.7\%) & 0.055$_{0.001}$( 16.8\%) & 0.750$_{0.003}$(  3.6\%) & 0.050 \\
	HierLE$_{\text{wl}}$ & \textbf{0.409$_{0.004}$(  6.0\%)} & \textbf{1.089$_{0.010}$( 16.1\%)} & \textbf{0.064$_{0.001}$( 19.5\%)} & \textbf{0.051$_{0.001}$( 23.0\%)} & \textbf{0.766$_{0.002}$(  5.8\%)} & 1.772 \\
	HierLE$_{\text{hinge}}$ & 0.424$_{0.004}$(  2.5\%) & 1.142$_{0.012}$( 12.0\%) & 0.067$_{0.001}$( 15.8\%) & 0.053$_{0.001}$( 19.4\%) & 0.755$_{0.003}$(  4.3\%) & 37.322 \\
	& \multicolumn{6}{c}{\textbf{CHTC} $k=3$} \\
	MSVM & 0.425$_{0.001}$(  0.0\%) & 1.464$_{0.004}$(  0.0\%) & 0.068$_{0.000}$(  0.0\%) & 0.061$_{0.000}$(  0.0\%) & 0.634$_{0.001}$(  0.0\%) & 551.649 \\
	SHSVM & 0.444$_{0.001}$( -4.3\%) & 1.506$_{0.004}$( -2.9\%) & 0.069$_{0.000}$( -1.9\%) & 0.063$_{0.000}$( -3.6\%) & 0.624$_{0.001}$( -1.7\%) & 432.623 \\
	SBSVM & 0.486$_{0.001}$(-14.2\%) & 1.668$_{0.004}$(-14.0\%) & 0.078$_{0.000}$(-14.0\%) & 0.071$_{0.000}$(-16.6\%) & 0.583$_{0.001}$( -8.1\%) & 900.615 \\
	HierCost & 0.335$_{0.001}$( 21.1\%) & 1.127$_{0.003}$( 23.0\%) & 0.052$_{0.000}$( 24.2\%) & 0.046$_{0.000}$( 23.4\%) & 0.718$_{0.001}$( 13.3\%) & 544.134 \\
	HierLE$_{\text{lin}}$ & 0.335$_{0.001}$( 21.4\%) & 1.100$_{0.003}$( 24.8\%) & 0.050$_{0.000}$( 27.0\%) & 0.045$_{0.000}$( 25.8\%) & 0.725$_{0.001}$( 14.3\%) & 0.548 \\
	HierLE$_{\text{wl}}$ & \textbf{0.324$_{0.001}$( 23.8\%)} & \textbf{1.059$_{0.003}$( 27.6\%)} & \textbf{0.048$_{0.000}$( 30.0\%)} & \textbf{0.043$_{0.000}$( 29.0\%)} & \textbf{0.735$_{0.001}$( 15.9\%)} & 20.461 \\
	& \multicolumn{6}{c}{\textbf{CHTC} $k=4$} \\
	MSVM & 0.499$_{0.001}$(  0.0\%) & 2.298$_{0.006}$(  0.0\%) & 0.068$_{0.000}$(  0.0\%) & 0.076$_{0.000}$(  0.0\%) & 0.603$_{0.001}$(  0.0\%) & 582.101 \\
	SHSVM & 0.521$_{0.001}$( -4.4\%) & 2.359$_{0.006}$( -2.6\%) & 0.068$_{0.000}$(  0.5\%) & 0.077$_{0.000}$( -1.7\%) & 0.594$_{0.001}$( -1.5\%) & 467.904 \\
	SBSVM & 0.565$_{0.001}$(-13.2\%) & 2.560$_{0.006}$(-11.4\%) & 0.076$_{0.000}$(-11.9\%) & 0.086$_{0.000}$(-12.7\%) & 0.556$_{0.001}$( -7.8\%) & 1027.250 \\
	HierCost & 0.412$_{0.001}$( 17.5\%) & 1.817$_{0.005}$( 20.9\%) & 0.052$_{0.000}$( 23.2\%) & 0.059$_{0.000}$( 21.9\%) & 0.686$_{0.001}$( 13.8\%) & 1202.397 \\
	HierLE$_{\text{lin}}$ & 0.417$_{0.001}$( 16.4\%) & 1.781$_{0.005}$( 22.5\%) & 0.049$_{0.000}$( 27.8\%) & 0.057$_{0.000}$( 25.6\%) & 0.693$_{0.001}$( 14.9\%) & 0.853 \\
	HierLE$_{\text{wl}}$ & \textbf{0.403$_{0.001}$( 19.3\%)} & \textbf{1.707$_{0.004}$( 25.8\%)} & \textbf{0.047$_{0.000}$( 31.4\%)} & \textbf{0.054$_{0.000}$( 29.3\%)} & \textbf{0.705$_{0.001}$( 17.0\%)} & 32.877 \\
	\bottomrule
	\end{tabular}
	\label{tab:CHTC}%
\end{sidewaystable}%

\section{Discussion}

In this paper, we propose an angle-based hierarchical classifier via exact label embedding.
In contrast to existing label embedding approaches, our embedding approach is an isometry map into a lower-dimensional space,   keeping    the hierarchy exactly and reducing  the complexity of the hypothesis space simultaneously.
Under the  (weighted) linear loss function, the solution is of a closed form, which makes it computationally efficient  for massive data.
Theoretical analyses show the advantages of the proposed method in the convergence rate over existing methods. Numerical experiments imply that our method performs excellently in both classification accuracy and computing, especially when the tree structure is complex and the sample size is large.

The idea of this paper can be extended in several aspects. First, we consider only the linear learner and the approach can be extended further into  kernel learning. Second, we consider the exact label embedding for  tree structure and single-labeled samples. Extending the  idea to  DAG structure or the multi-labeled case is an interesting future research topic. Third, when the number of features is large, some sparse penalty functions such as the $l_1$ penalty can be used to select features.

\bibliographystyle{Chicago}
\bibliography{reference}
\end{document}